\documentclass[11pt,a4paper]{article}

\usepackage{amsmath,amssymb}
\usepackage{epsfig,graphicx}
\usepackage{subfigure}
\usepackage{graphicx}
\usepackage{rotating}
\usepackage{cancel}
\usepackage{bm}
\usepackage{color}
\usepackage{comment}
\usepackage{cite}
\usepackage{psfrag}
\usepackage[T1]{fontenc}
\usepackage[utf8]{inputenc}
\usepackage{footmisc}
\usepackage{authblk}
\usepackage{hyperref}

\def\bra{\langle}
\def\ket{\rangle}
\def\beq{\begin{equation}}
\def\eeq{\end{equation}}
\newcommand{\C}[1]{\mathcal{#1}}

\def\occupancy{A_0}

\begin{document}
\numberwithin{equation}{section}
\title{Attractive vs.~repulsive interactions in the Bose-Einstein condensation dynamics of relativistic field theories
\vspace{2.5cm}
\Large{\textbf{\vspace{0.5cm}}}}

\author[1]{J. Berges}
\author[2]{K. Boguslavski}
\author[1]{A. Chatrchyan\thanks{chatrchyan@thphys.uni-heidelberg.de}}
\author[1]{J. Jaeckel}
\affil[1]{\small{\em Institut f\"ur theoretische Physik, Universit\"at Heidelberg,} \authorcr \small{\em Philosophenweg 16, 69120 Heidelberg, Germany}}
\affil[2]{\small{\em Department of Physics, University of Jyv\"{a}skyl\"{a},} \authorcr \small {\em P.O.~Box 35, 40014 University of Jyv\"{a}skyl\"{a}, Finland}}

\renewcommand\Authands{ and }

\date{}
\maketitle


\begin{abstract}
\noindent

We study the impact of attractive self-interactions on the nonequilibrium dynamics of relativistic quantum fields with large occupancies at low momenta. Our primary focus is on Bose-Einstein condensation and nonthermal fixed points in such systems. As a model system we consider $\C{O}(N)$-symmetric scalar field theories. We use classical-statistical real-time simulations, as well as a systematic $1/N$ expansion of the quantum (2PI) effective action to next-to-leading order. When the mean self-interactions are repulsive, condensation occurs as a consequence of a universal inverse particle cascade to the zero-momentum mode with self-similar scaling behavior. For attractive mean self-interactions the inverse cascade is absent and the particle annihilation rate is enhanced compared to the repulsive case, which counteracts the formation of coherent field configurations. For $N \geq 2$, the presence of a nonvanishing conserved charge can suppress number changing processes and lead to the formation of stable localized charge clumps, i.e.~$Q$-balls. 

\end{abstract}

\newpage


\section{Introduction}

Nonequilibrium quantum fields with large occupation numbers at low momenta are frequently encountered in the context of cosmology. Important examples include the decay of coherent oscillations of the inflaton during the reheating stage after inflation \cite{Kofman:1994rk}, or the production of dark matter axions from the misalignment mechanism and from the decay of axionic strings and domain walls~\cite{Preskill:1982cy,Abbott:1982af,Dine:1982ah, Davis:1986xc,Lyth:1991bb}. Therefore, understanding the dynamics of such systems is important. Particularly interesting is the possibility of Bose-Einstein condensation in such systems, which is the macroscopic occupation of the quantum state with the lowest energy. In the case of axions the resulting collective quantum behavior may have important observational consequences in cosmology (see e.g.~\cite{Sikivie:2012gi,Banik:2013rxa, Sikivie:2013joa}) and would leave distinct imprints in direct detection experiments~\cite{Duffy:2006aa,Irastorza:2012jq,Jaeckel:2013sqa}. While many arguments have focussed on the formation rate of a condensate~\cite{Sikivie:2009qn,Erken:2011vv,Erken:2011dz,Saikawa:2012uk,Davidson:2013aba,Noumi:2013zga,Berges:2014xea}, a qualitatively even more important point is the attractive nature of the relevant interactions~\cite{Davidson:2014hfa,Guth:2014hsa}
that tends to favor localised structures instead of a spatially constant condensate~\cite{Guth:2014hsa}. 
Inspired by this it is one of our main aims to study the impact of attractive interactions and to delineate the differences to the repulsive case.

In recent years there has been a significant advance in the theoretical understanding of the dynamics of isolated highly occupied quantum fields. Many characteristic properties of the dynamics in such extreme conditions turn out to be insensitive to details of the underlying model and initial conditions. This allows one to classify theories with different microscopic descriptions into universality classes \cite{Berges:2014bba, Orioli:2015dxa}. This notion of universality is based on the existence of nonthermal fixed points \cite{Berges:2008wm, Berges:2008sr} that are nonequilibirium attractor solutions with self-similar scaling behavior of correlation functions and of the particle momentum distribution. Self-similarity in this case is associated to the transport of some conserved quantity in momentum space \cite{Zakharov:1985ki, Micha:2004bv}. Self-similar scaling regions can represent transport of energy towards high momenta, or particle number transport towards the zero-mode. While the first case drives the thermalization process by pushing the typical hard scale to larger momenta \cite{Micha:2004bv}, the second one may lead to the formation of a Bose-Einstein condensate out-of-equilibrium \cite{Berges:2012us}. Both cascades have been observed in different regions of the same momentum distribution function in scalar models with positive quartic self-coupling \cite{Orioli:2015dxa, Berges:2015ixa}.

In general, for scalar $N$-component fields $\varphi_a$, $a=1,...,N$, with self-interactions of the form $\sim \lambda_{2n} (\varphi_a \varphi_a)^n$, positive or negative signs of the couplings determine whether they are repulsive or attractive. While repulsive interactions have the tendency to dilute concentrations of energy density over position space, attractive interactions encourage such local concentrations. It is thus not surprising that condensation dynamics is strongly affected by the presence of attractive interactions, since the lowest-energy configurations in this case are localized ``clumps''. Therefore, it is an interesting question how characteristic features of the dynamics depend on the type of self-interactions. 

In the present work we investigate the influence of attractive self-interactions on the far-from-equilibrium dynamics of scalar fields and, in particular, their impact on Bose-Einstein condensation. As our model we consider $\C{O}(N)$-symmetric relativistic field theory\footnote{While the $N=1$ case can be viewed as a toy model for axions, multi-component fields are, for example, encountered in models of hybrid inflation \cite{Linde:1993cn}, Higgs inflation \cite{Bezrukov:2007ep}.} in 3+1 dimensions, exhibiting an attractive quartic self-interaction ($\lambda<0$) and stabilized by a repulsive sextic term, so that the potential has an $\C{O}(N)$-symmetric global minimum. We are interested in the nonperturbative regime of large occupancies $f \sim 1/|\lambda|$ for typical momenta, in the weak coupling regime with $|\lambda| \ll 1$. Our analysis is based on real-time classical-statistical lattice simulations, using the possibility of mapping the dynamics of highly occupied quantum fields onto a classical-statistical field theory evolution\footnote{The range of validity of classical-statistical approximation for the description of the underlying quantum theory has been studied in detail for similar models with $\lambda \varphi^4$ interaction \cite{Berges:2013lsa}. For given regularization of the field theory it requires sufficiently high typical occupancies for small enough coupling. Corresponding restrictions apply to the size of the $\varphi^6$ coupling, which is taken into account in our study by choosing $\lambda_6 \sim \lambda^2/m^2$, with $m^2 \sim {\mathrm{p}_0}^2$, as defined in Sec.~\ref{sec:model}.} \cite{Khlebnikov:1996mc, Aarts:2000mg}. In addition to this, we develop a vertex-resummed kinetic theory, based on the $1/N$ expansion of the quantum 2PI effective action in the presence of quartic and sextic interactions to next-to-leading order \cite{Berges:2001fi}. This extends well-established kinetic descriptions \cite{Micha:2004bv, Zakharov:1985ki} to the considered highly occupied regime, in analogy to the repulsive $\varphi^4$ theory \cite{Berges:2010ez, Orioli:2015dxa}. We derive the corresponding effective kinetic equation, discuss its properties and compare with our observations from classical-statistical simulations.

The outline of this work is as follows. In Sec.~\ref{sec:model} we describe our model and the considered initial conditions. In Sec.~\ref{sec:dynamics_N1} we present the results from the classical-statistical simulations for the single-component model. In Sec.~\ref{sec:dynamics_N_comp} we extend that discussion to the case of multi-component fields and derive the vertex-resummed kinetic description. We conclude in Sec.~\ref{sec:conclusion}. Throughout this paper we work in Minkowski spacetime with metric $\eta_{\mu\nu}=(1,-1,-1,-1)$ and set $\hbar=k_B=c=1$.


\section{The model}
\label{sec:model}

In this paper we study $\C{O}(N)$-symmetric scalar field theories in flat, 3+1 dimensional spacetime. We consider potentials that contain an attractive self-interaction, but have a global minimum at $\varphi = 0$. The simplest form of a potential capturing these properties is given by
\beq
  U( \varphi )\: = \:   {\frac{m^{2}}{2}} \, \varphi_a(x) \varphi_a(x) \: + \: {\frac {\lambda}{4!N}} \, (\varphi_{a}(x) \varphi_{a}(x))^{2} \: + \: {\frac {\lambda_6}{6!N^{2}}}\,  (\varphi_{a}(x) \varphi_{a}(x))^{3}.
\label{eq:potential}
\eeq
Here summation over repeated indices is implied and the factors of $1/N$ are chosen such that the classical action $S[\varphi]= \int_x[ \frac{1}{2}\partial^\mu \varphi_a\partial_\mu \varphi_a- U ( \varphi ) ]$ scales proportional to $N$. The parameters $m^2$ and $\lambda_6$ are positive. For completeness we consider both cases of $\lambda>0$ and $\lambda<0$, and refer to those models throughout this paper as repulsive and attractive, respectively. We assume weak couplings, in particular $|\lambda| \ll 1$.

\begin{figure}[tp!]
  \centering
    \includegraphics[scale=0.93]{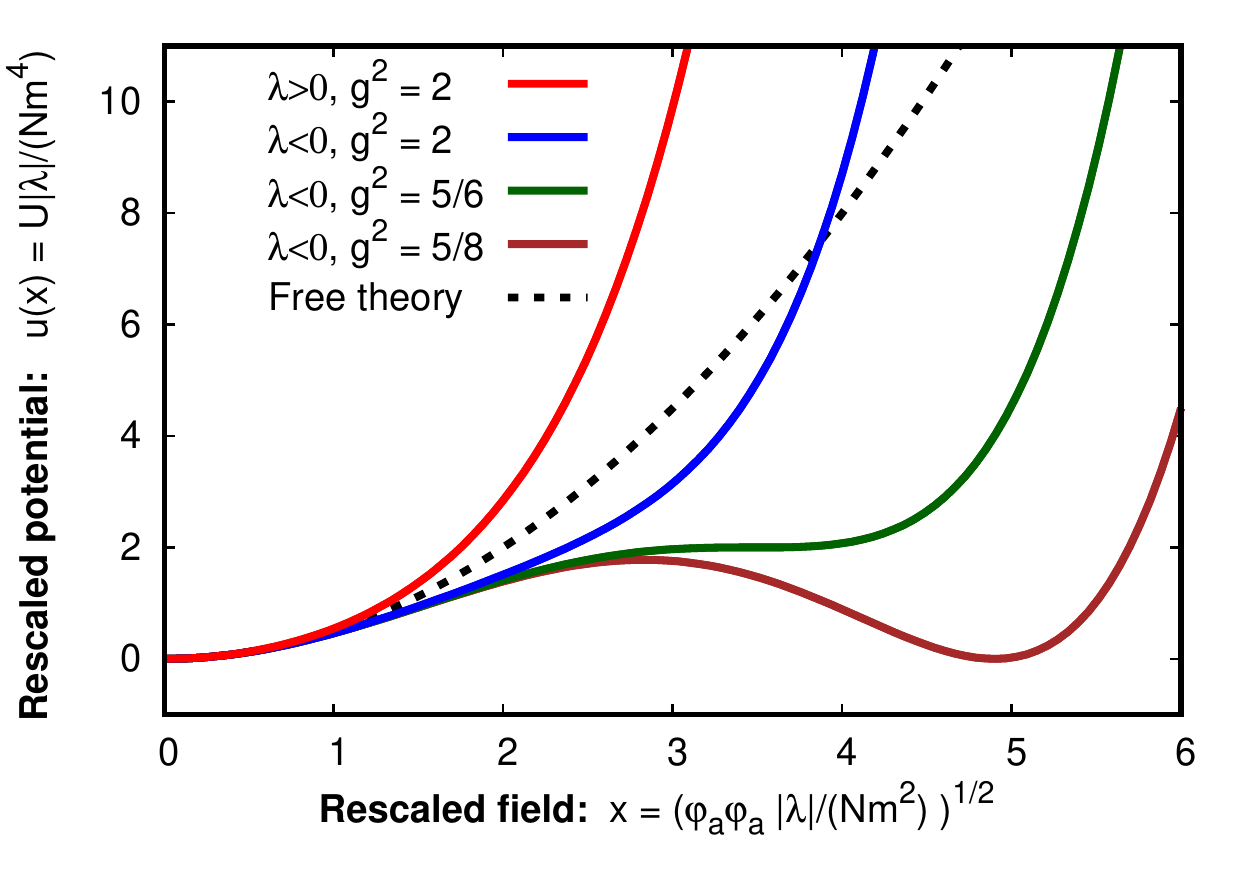}
\caption{The rescaled potential of Eq.~(\ref{eq:rescaled_potential}) as a function of the rescaled field amplitude for different signs of $\lambda$ and several values of $g^2$. The black dashed line corresponds to the quadratic potential of the free theory, $u(x)=x^2/2$. Roughly speaking, the theory is attractive when the potential lies below that of the free theory.}
\label{fig:potential}
\end{figure}

After the rescaling prescriptions $x=\sqrt{\frac{|\lambda|}{m^2}\frac{\varphi_a \varphi_a}{N}}$ and $u(x)=\frac{|\lambda|}{ Nm^4}U(\varphi)$ the potential can be written as
\beq
 u(x)  \: = \: \frac{x^2}{2} \: + \: \mathrm{sgn}(\lambda) \frac{x^4}{4!} \: + \: g^2\; \frac{x^6}{6!}\,.
 \label{eq:rescaled_potential}
\eeq
The dimensionless parameter $g^2$, defined as
\beq
g^2= \frac{ \lambda_6 m^2 }{ \lambda^2 }\,,
\eeq
characterizes the strength of the sextic coupling. In the case $\lambda<0$, the potential falls below quadratic near its minimum, but stays positive everywhere if $g^2$ is sufficiently large ($g^2 > 5/8$). Similarly, it becomes convex for $g^2 > 3/2$. This is demonstrated in Fig.~\ref{fig:potential}, where the rescaled potential is shown for several values of $g^2$, together with the free theory parabola $u(x)=x^2/2$, which may be viewed as a dividing line between repulsion and attraction.

We consider spatially homogeneous and isotropic quantum systems, and employ Gaussian initial conditions, which can be formulated in terms of the macroscopic field\footnote{The brackets $\bra...\ket$ indicate the density matrix average.} $\phi_a(t)= \bra \hat \varphi_a(t,\mathbf{x}) \ket$ and the single-particle distribution function $f(t,|\mathbf{p}|)$ at the initial time $t=0$ \cite{Berges:2015kfa}. In order to observe the build-up of correlations and condensation, we start with a vanishing macroscopic field and large initial occupation numbers $f \gtrsim 1/|\lambda|$ of the following form
\beq
f(0,|\mathbf{p}|)=\frac{\occupancy}{ |\lambda| }\; \Theta(\mathrm{p}_0-|\mathbf{p}|),\qquad \qquad \phi(0)=\dot \phi(0)=0\,.
\label{eq:box}
\eeq
These initial conditions correspond to a `box' in momentum space up to the momentum scale $\mathrm{p}_0$, with the amplitude $\occupancy \gtrsim 1$. Another type of initial conditions that is also often employed in the literature corresponds to a large macroscopic field $\phi(0) \sim 1/\sqrt{ |\lambda|}$ in the absence of initial occupation numbers with $f(0,|\mathbf{p}|)=0$ \cite{Micha:2004bv}. However, in a transient regime of parametric resonance, the evolution of the field $\phi$ triggers growth of unstable modes and the system eventually becomes highly occupied, recovering the fluctuation initial conditions \cite{Berges:2015ixa, Micha:2004bv}. We will thus only employ box-type initial conditions (\ref{eq:box}).

Despite starting with Gaussian initial conditions, self-interactions render the dynamics of the system non-Gaussian and higher-order non-factorizable correlation functions build up for $t>0$. Although the notion of particles is not uniquely defined in interacting relativistic field theory, the following definition of $f(t,|\mathbf{p}|)$ turns out to provide a useful quasi-particle interpretation for scalar systems \cite{Berges:2015kfa, Berges:2008wm}. With the help of the anticommutator expectation value
\beq
\label{eq:stat_prop}
F(t,t',\mathbf{\Delta x}) \: =\: \frac{1}{2N} \bra \{ \hat{\varphi}_{a}(t,\mathbf{x}),  \hat{\varphi}_{a}(t',\mathbf{x'}) \}\ket,
\eeq
where\footnote{Due to the system's homogeneity, the Fourier transformation is performed with respect to the relative coordinates: $F(t,t',\mathbf{p}) = \int d^3\mathbf{\Delta x} \exp(-i \mathbf{p \mathbf{\Delta x}})\, F(t,t',\mathbf{\Delta x})$.} $\mathbf{\Delta x=x-x'}$, it can be expressed as\footnote{A closely related alternative definition of the distribution function, also employed in the literature  \cite{Berges:2012us, Orioli:2015dxa}, is $\Bigl(f(t,|\mathbf{p}|)+\frac{1}{2}+(2\pi)^3\delta^{(3)}(\mathbf{p})n_{\mathrm{cond}}(t)\Bigr)/\omega(t,\mathbf{p})=F(t,t,\mathbf{p})$. It turns out to yield the same distribution function, if a relativistic form for $\omega$ is employed with a suitably chosen effective mass.\label{fn:definition_distrib}}
\beq
\label{eq:distr}
f(t,|\mathbf{p}|)+\frac{1}{2}+(2\pi)^3\delta^{(3)}(\mathbf{p})\;n_{\mathrm{cond}}(t)=\sqrt {F(t,t',\mathbf{p})\;\partial_t \partial_{t'} F(t,t',\mathbf{p})}\Big|_{t'=t}\,.
\eeq
We have separated a term proportional to the $\delta$-function to account for a possible condensation into the zero-momentum mode. Similarly, we define the quasi-particle frequency
\beq
\label{eq:omega}
\omega(t,\mathbf{p})=\sqrt {\frac{\partial_t \partial_{t'} F(t,t',\mathbf{p})}{F(t,t',\mathbf{p})}}\Bigg|_{t'=t},
\eeq
which can be compared to a relativistic dispersion relation $\omega(t,\mathbf{p})=\sqrt{\mathbf{p}^2+M^2(t)}$ with an effective mass $M(t)$.


\section{Dynamics for single-component model ($N=1$)}
\label{sec:dynamics_N1}

We start our discussion with the simplest case of a single-component field theory. Since we consider systems with large occupation numbers at characteristic momenta, perturbative approaches based on a coupling expansion are inapplicable. On the other hand, the large occupation numbers of typical modes that dominate the energy density strongly exceed unity, such that genuine quantum effects become subdominant and the dynamics of the system becomes essentially classical. More precisely, in this regime quantum field theory can be mapped onto a classical-statistical field theory \cite{Khlebnikov:1996mc, Aarts:2000mg}, which can be simulated numerically. Otherwise, as soon as typical occupation numbers become of order unity, $f \lesssim 1$, the mapping becomes inaccurate \cite{Jeon:2004dh, Berges:2013lsa, Epelbaum:2014yja, Epelbaum:2014mfa} and the system leaves the classical regime thermalizing eventually \cite{Berges:2016nru}. Thus, we restrict ourselves to the case of sufficiently large occupation numbers and study the scalar systems with classical-statistical simulations in this section. 

In such lattice simulations one samples over the initial conditions and evolves each realization according to the classical field equation of motion. Correlation functions are obtained by taking ensemble averages over the classical trajectories, where quantum observables are translated to their classical counterparts. For instance, the anticommutator expectation value $\bra \frac{1}{2}\{.,.\} \ket$ in (\ref{eq:stat_prop}) is replaced by the averaged product of fields in the classical-statistical framework. 

Numerical simulations are performed on a 3D cubic lattice with periodic boundary conditions with up to $1024^3$ points and lattice spacings in the range of $0.0625$ - $0.5$ $\mathrm{p_0^{-1}}$, with $\mathrm{p}_0$ introduced by the initial conditions (\ref{eq:box}).\footnote{In the central and lower plots in the left panel of Fig.~2 we combined two data sets with different lattice spacings ($0.0625\,\mathrm{p_0^{-1}}$ with $0.25\,\mathrm{p_0^{-1}}$, and $0.1\,\mathrm{p_0^{-1}}$ with $0.25\,\mathrm{p_0^{-1}}$, respectively) to show a larger momentum range of the distribution function. In the overlapping momentum region, the two simulations agree to very good accuracy.} The classical equations of motion are solved using the standard leap-frog algorithm and the details are presented in appendix A. The distribution function $f(t,\mathbf{p})$ and the dispersion relation $\omega(t,\mathbf{p})$ are computed by use of Eqs.~(\ref{eq:distr}) and (\ref{eq:omega}), respectively. The latter is also used to calculate the effective mass $M(t)$ by fitting $\sqrt{\mathbf{p}^2+M^2(t)}$. Another important quantity is the total particle number density\footnote{On the lattice, integrals are translated to sums over momenta according to $\int \frac{d^3\mathbf{p}}{(2\pi)^3} \rightarrow \frac{1}{V}\sum_\mathbf{p}\,$.}
\beq
\label{eq:tpnd}
n(t)= \int \frac{d^3\mathbf{p}}{(2\pi)^3}\: f(t,|\mathbf{p}|) + n_{\mathrm{cond}}(t).
\eeq
If not stated otherwise, we employ $g^2 = 2$ and $m = 0.7\, \mathrm{p}_0$. More generally, all dimensionful quantities shown in the plots of this section are made dimensionless by rescaling with appropriate powers of $\mathrm{p}_0$. In addition, we plot rescaled functions $F \mapsto |\lambda|F$, $f \mapsto |\lambda|f$ and $n \mapsto |\lambda| n$, since for fixed $m^2$, $\occupancy$ and $g^2$, the classical-statistical dynamics of these combinations does not depend on the absolute value of $\lambda$. However, the dynamics crucially depends on its sign. Both cases of $\lambda>0$ and $\lambda<0$ are considered separately.


\subsection{Repulsive ($\lambda>0$)}
\label{sec:N1_repulsive}

For a better understanding of the role of the negative coupling, we first briefly discuss the dynamics for positive $\lambda$, which corresponds to repulsive self-interactions. The same model without the sextic coupling has been extensively studied previously \cite{Micha:2004bv, Berges:2012us, Orioli:2015dxa}.  For the range of initial occupancies we consider, we find that the presence of the sextic coupling does not introduce any significant difference for the dynamics.

For the numerical simulations we employ ``box'' initial conditions with $\occupancy$ in the range $20-500$. In the beginning of the time evolution the momentum distribution transforms from the initial ``box'' into a smoother function, exhibiting a dual cascade. The direct UV cascade populates the initially empty high-momentum modes. At the same time, the inverse cascade at low momenta constantly transports particles to the zero-momentum mode, leading to far-from-equilibrium Bose-Einstein condensation \cite{Berges:2012us}. After some transient time the dynamics of each of the cascades becomes self-similar:
\beq
f(t,|\mathbf{p}|)=t^{\alpha}f_S(t^\beta |\mathbf{p}|)\,,
\label{eq:self_similar}
\eeq
with separate sets of real scaling exponents $\alpha$ and $\beta$ and fixed point distributions $f_S$. This self-similar dynamics reflects the system being in the vicinity of a nonthermal fixed point \cite{Berges:2008wm, Berges:2008sr}.

\begin{figure}[tp!]
  \centering
  	\includegraphics[scale=0.82]{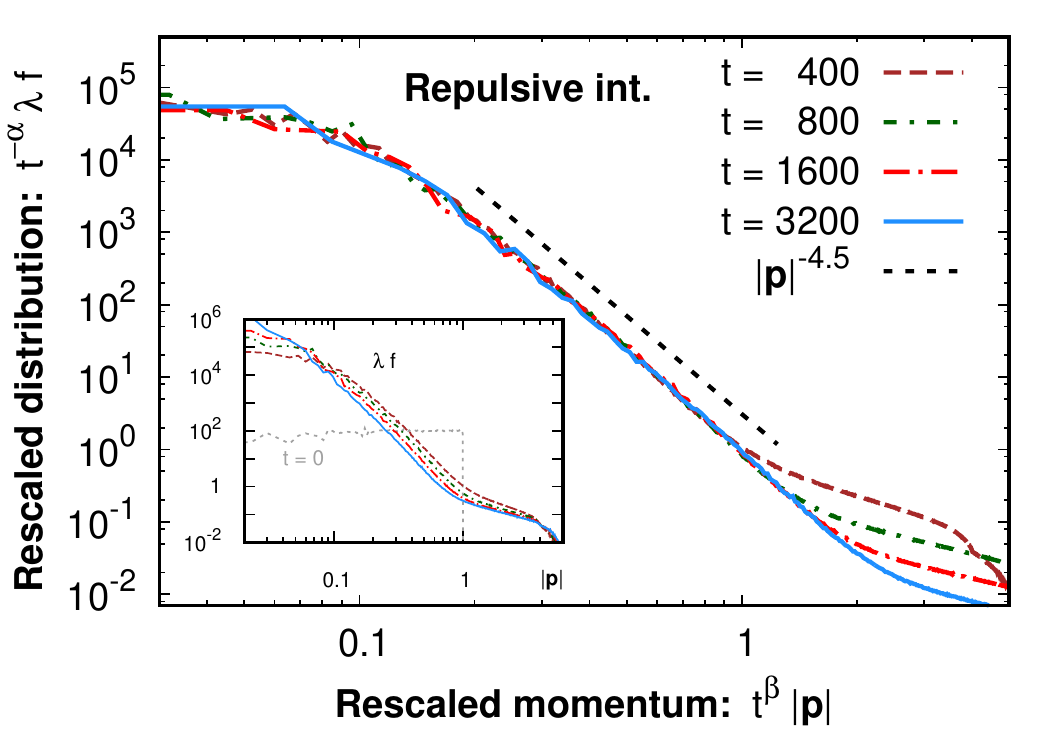}
    \includegraphics[scale=0.82]{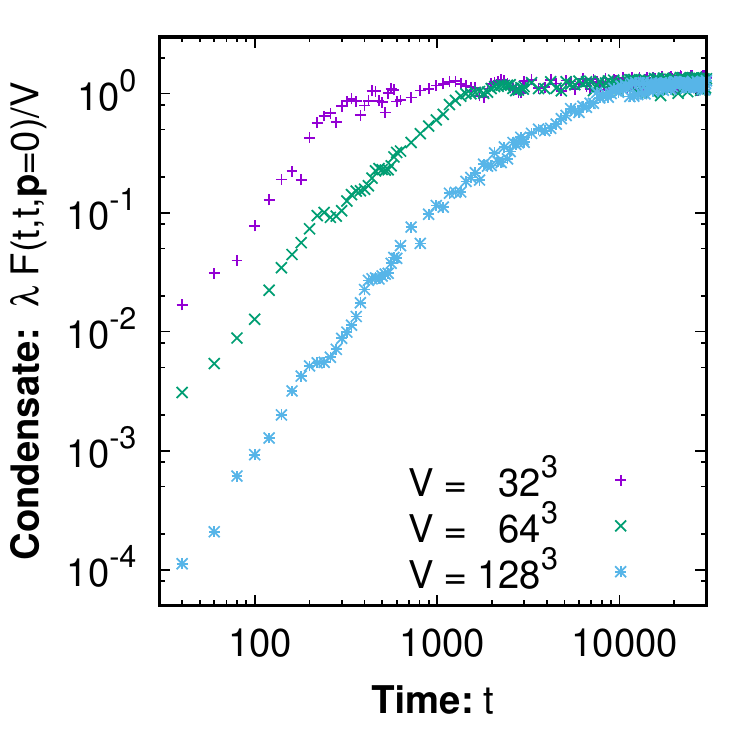}
    \includegraphics[scale=0.82]{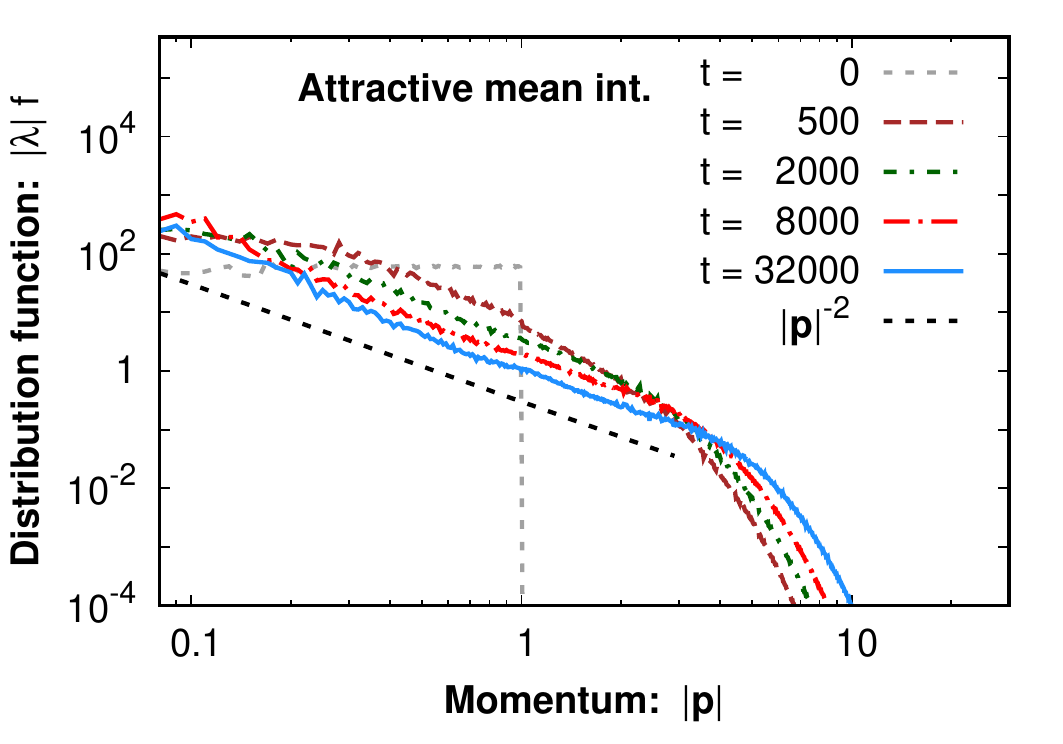}
    \includegraphics[scale=0.82]{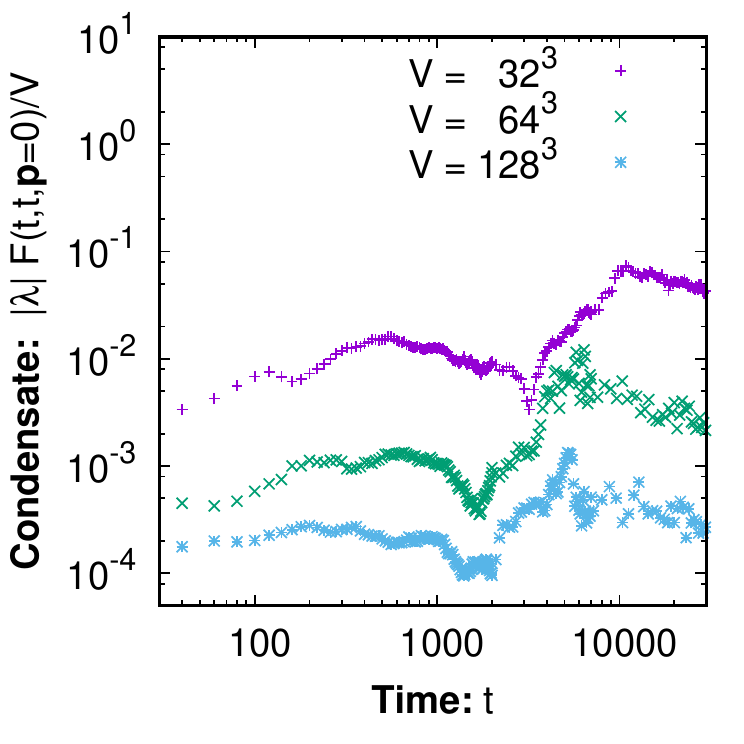}
    \includegraphics[scale=0.82]{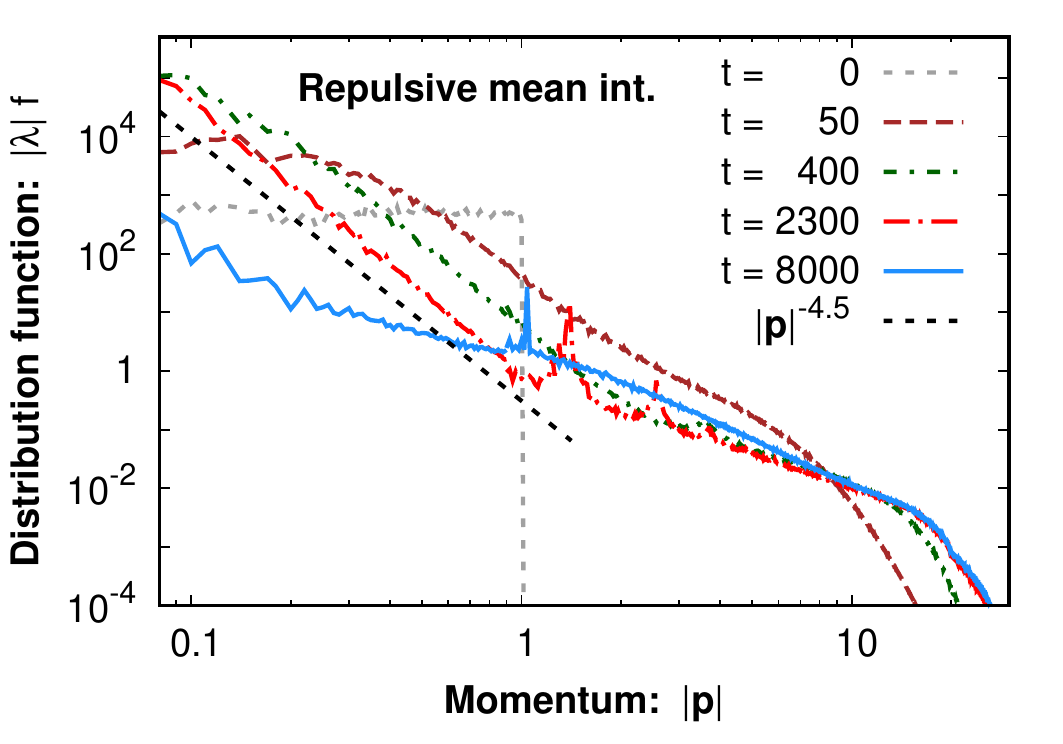}
    \includegraphics[scale=0.82]{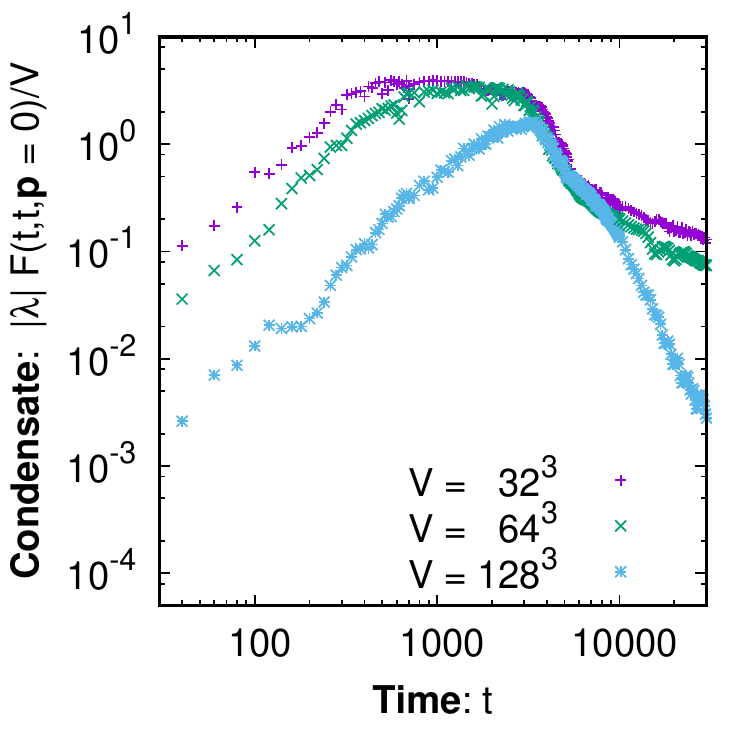}
\caption{Snapshots of the distribution function as a function of the momentum (left column) and the zero-momentum correlation function divided by the volume for several volumes (right column). {\bf Upper row:} Repulsive theory ($\lambda > 0$, $\occupancy = 100$); distribution function and momentum in the left panel are rescaled with powers of time, with the inset showing the original distribution without any rescaling; right panel demonstrates formation of a condensate. {\bf Central, lower rows:} Attractive theories ($\lambda < 0$) with attractive ($\occupancy = 60$) and repulsive ($\occupancy = 500$) mean interactions initially. Note that the momentum is not rescaled.}
\label{fig:rep-attractive}
\end{figure}

In the upper left panel of Fig.~\ref{fig:rep-attractive} we show the rescaled distribution function $t^{-\alpha}f(t,|\mathbf{p}|)$ as a function of the rescaled momentum $t^\beta |\mathbf{p}|$ at different times for the values $\alpha = 3/2$ and $\beta = 1/2$ while the inset gives the same distributions without rescaling. Since the rescaled curves lie on top of each other, the system follows a self-similar evolution with the given values for the exponents $\alpha$ and $\beta$. The fixed point distribution $f_S$ exhibits an approximate power-law behavior $\sim |\mathbf{p}|^{-\kappa}$, with $\kappa \approx 4.5$. The same values for these three exponents have been obtained also for the $\varphi^4$ model \cite{Orioli:2015dxa}. 

The occupancy of the zero-momentum mode grows during the self-similar evolution approximately as a power-law $\sim t^{\alpha}$. This can be directly understood from (\ref{eq:self_similar}), by setting $\mathbf{p}=0$. For a finite volume this growth continues until $F(t,t,\mathbf{p}=0)$ becomes proportional to the volume $V$, which signals the emergence of the condensate according to \cite{Berges:2012us} (see also footnote \ref{fn:definition_distrib}) with the replacement $(2\pi)^3\delta(\mathbf{0}) \rightarrow V$. This is demonstrated in the upper right panel of Fig.~\ref{fig:rep-attractive}, where the quantity $F(t,t,\mathbf{p}=0)/V$ is shown as a function of time for several volumes. All of this has also been observed in the $\varphi^4$ model \cite{Orioli:2015dxa}. There it was also shown that this dynamics implies that the time required for creating a condensate scales with volume.

The inverse cascade represents the transport of conserved particle number to the zero mode. Particle number conservation follows from the relation $\alpha=3\beta$, which was discussed above:
\beq
\int_{ t^{-\beta}\mathrm{p_1}}^{ t^{-\beta}\mathrm{p_2}} \frac{d^3 \mathbf{p}}{(2\pi)^3}f(t,\mathbf{p})=t^{\alpha-3\beta} \int_{\mathrm{p_1}}^{\mathrm{p_2}} \frac{d^3\mathbf{p}}{(2\pi)^3}f_S(\mathbf{p})=\mathrm{const}.
\label{eq_icpnc}
\eeq
Moreover, the condensate stays approximately constant after its creation, as can be seen in the upper right panel of Fig.~\ref{fig:rep-attractive}. Hence, also the total particle number density $n(t)$ is approximately conserved, since its dominant contribution comes from the infrared modes and the condensate. Therefore, at this stage of the evolution, elastic scatterings are the dominant processes for the dynamics in repulsive scalar systems and lead to the emergence of a long-lived condensate\footnote{The decay of the condensate and the infrared modes has been studied in \cite{Moore:2015adu}. There it was shown that with an increasing mass parameter $m$ such decays become more suppressed. For the values of the mass parameter and the initial occupancy considered here we confirm that the total particle number density is practically conserved.}.

At high momenta, another self-similar scaling region emerges after some time. There we have extracted scaling exponents $\beta'  \approx -1/4.2$  and $\alpha' \approx 4 \beta'$, with error estimates of the order of 10\%. This ratio of the exponents corresponds to a transport of conserved energy towards high momenta in the ultrarelativistic regime with $\omega_\mathbf{p} \approx |\mathbf{p}|$. We have also verified that these exponents are unaffected by the presence of the sextic coupling\footnote{The scaling exponents of the self-similar energy cascade have been studied in \cite{Micha:2004bv}, and for $k \leftrightarrow l$ scattering processes they are given by $\beta'=-1/(2(k+l)-1)$. Our extracted values slightly exceed $\beta'=-1/5$, which would correspond to $2 \leftrightarrow 1 + \textrm{soft}$ scatterings, where the soft particle corresponds to the involvement of the condensate \cite{Micha:2004bv}}. 

The scaling exponents of both cascades thus seem to be independent of the microscopic details of the model. In Sec.~\ref{sec:dynamics_N_comp} we explain the reason for this from the point of view of the 2PI $1/N$ expansion to NLO.

\subsection{Attractive ($\lambda<0$)}
\label{sec:N1_attractive}

We now consider the attractive model with $\lambda<0$. In addition, we take the positive sextic coupling characterized by $g^2$ sufficiently large such that the potential has its global minimum at the origin. Results are presented for $g^2=2$.

The strength of the fluctuations, characterized by the parameter $\occupancy$, determines which of the two self-interactions is more important for the dynamics. 
Therefore, we split our discussion into two parts\footnote{Qualitatively the two regimes can be easily understood from Fig.~\ref{fig:potential}, where the blue potential curve lies below the free theory curve for small field values and above it for large field values. We will quantify this for the case of an $N$ component theory in Sec.~\ref{sec:dynamics_N_comp}, Eq.~\eqref{eq:lmod}.}
. We first consider the case of relatively weak initial fluctuations, such that the negative, attractive quartic coupling is more important and mean interactions are attractive. In the second regime, initial fluctuations are strong and the positive repulsive sextic coupling dominates, turning the mean interactions repulsive. In systems with high occupation numbers we will even encounter a transition between these regimes during the time evolution.

\subsubsection{Attractive mean interactions}

The inverse particle cascade is absent in this case. This is demonstrated for $\occupancy = 60$ in the central left panel of Fig.~\ref{fig:rep-attractive}, which contains several snapshots of the distribution function. While a transient growth of the low momentum occupancies is observed at early times, it does not lead to an enhancement of infrared modes as strong as in the repulsive models, $\sim |\mathbf{p}|^{-4.5}$, that were discussed above. For comparison, a power law $\sim |\mathbf{p}|^{-2}$ is additionally shown in the plot.

Due to the absence of an inverse particle cascade, the zero-momentum mode does not develop a condensate part. This is demonstrated in the central right panel of Fig.~\ref{fig:rep-attractive}, where the time evolution of the correlation function $F(t,t,\mathbf{p}=0)$ divided by the volume $V$ is shown for several volumes. As was mentioned in the previous section, for a finite volume, the presence of a condensate would lead to a contribution in $F(t,t,\mathbf{p}=0)$, which scales proportional to $V$. Such a contribution is clearly absent in the attractive case, which becomes more stringent when compared to the repulsive system depicted in the upper right panel of Fig.~\ref{fig:rep-attractive}. In other words, no long-range order is being established. 

At hard momenta, a self-similar cascade develops after some transient time. The extracted values of the scaling exponents $\beta' \approx -1/5$ and $\alpha' \approx 4\beta'$ are slightly smaller than those observed in the repulsive model\footnote{In the absence of a condensate one would expect elastic $2 \leftrightarrow 2$ scatterings to be the dominant processes for the direct cascade at high momenta, which would lead to $\beta' = -1/7$ \cite{Micha:2004bv}. As in the repulsive model, we observe some discrepancy from the expected value of $\beta'$. Nevertheless, the hierarchy of the two $\beta'$ exponents, that $|\beta'|$ for the repulsive model exceeds its value in the attractive system, qualitatively coincides with expectations.}. As in the repulsive theory, the ratio $\alpha' / \beta' = 4$ is characteristic of a transport of conserved energy density towards high momenta. Hence, the dynamics at high momenta can be also interpreted as a direct energy cascade. 

As compared to repulsive systems, number changing processes appear to be more efficient in the attractive model. The time evolution of particle number density $n(t)$ is shown in the upper left panel of Fig.~\ref{fig:transition}, where the lines with $\occupancy = 12$ and $\occupancy = 60$ correspond to the considered regime with mean attractive interactions. After some transient time $n(t)$ decreases approximately as a power law $n(t) \sim t^{\beta'}$. This exponent can be understood by plugging the self-similar ansatz into Eq.~(\ref{eq_icpnc}) in the absence of a highly occupied infrared region. 
This loss of particles is quite different from the repulsive model discussed above. The repulsive model features an enhanced infrared region, and its total particle number density is approximately conserved, despite the presence of a direct energy cascade with $\alpha' = 4\beta'$ at hard momenta.  

The absence of long-range order in our considered system can be viewed as a manifestation of the tendency of attractive self-interactions to fragment the field into localized configurations \cite{Guth:2014hsa}. Importantly, number changing processes prevent the formation of visible spatial structures.\footnote{Although in single simulations relatively long-lived configurations with energy density concentration may be observed, they do not affect the distribution function in any significant way. We have verified this by removing them by explicitly replacing the field value in these regions by a small fixed field value and comparing the distribution function afterwards. We note that the importance of such configurations may depend on the chosen parameter region, see for example \cite{Amin:2010dc}.}

\subsubsection{Repulsive mean interactions}

If the initial occupancy $\occupancy$ is sufficiently increased, with other parameters fixed, the repulsive sextic term becomes more important initially. In this case an early stage is observed, during which the distribution function exhibits a dual cascade. This is shown in the lower left panel of Fig.~\ref{fig:rep-attractive}, where we show several snapshots of the distribution function at different times. At early times, a self-similar inverse particle cascade is observed with the same scaling exponents $\alpha \approx 3/2$, $\beta \approx 1/2$ and an approximate power-law behavior $\sim |\mathbf{p}|^{-4.5}$ as in the repulsive $\lambda>0$ case. However, at later times the dynamics starts to deviate from a self-similar evolution and eventually, the infrared enhancement decays. 

A similar evolution is observed for the zero-momentum mode. At early times, condensation occurs as in the repulsive model (especially for sufficiently small volumes, where the condensation time is small enough), as is visible in the lower right panel of Fig.~\ref{fig:rep-attractive}. However, at later times, the condensate collapses. 

\begin{figure}[t!]
  \centering
  	\includegraphics[scale=0.835]{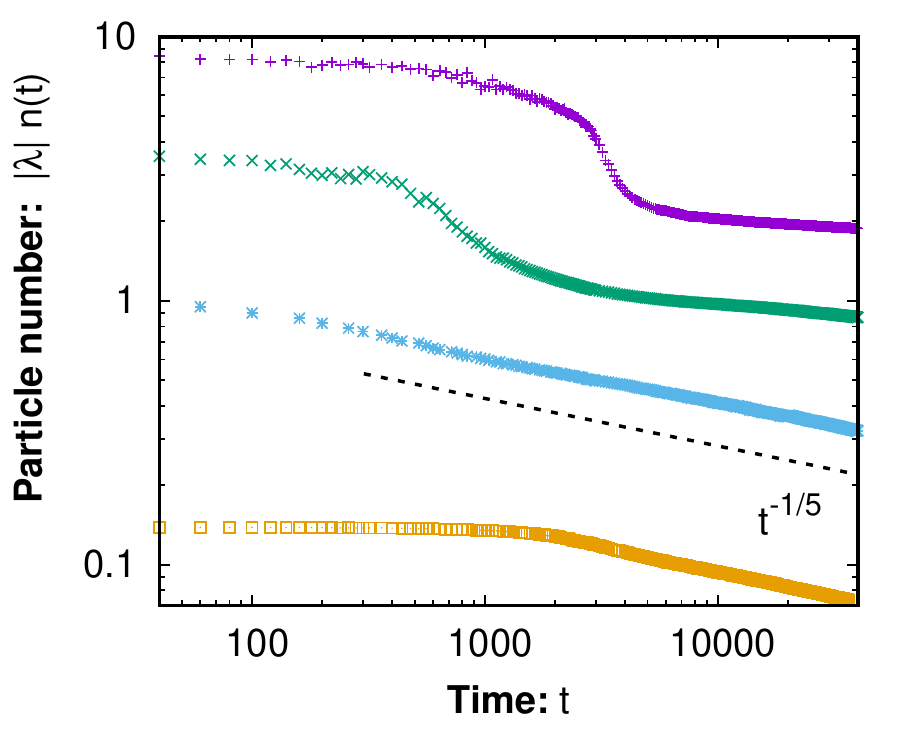}
    \includegraphics[scale=0.835]{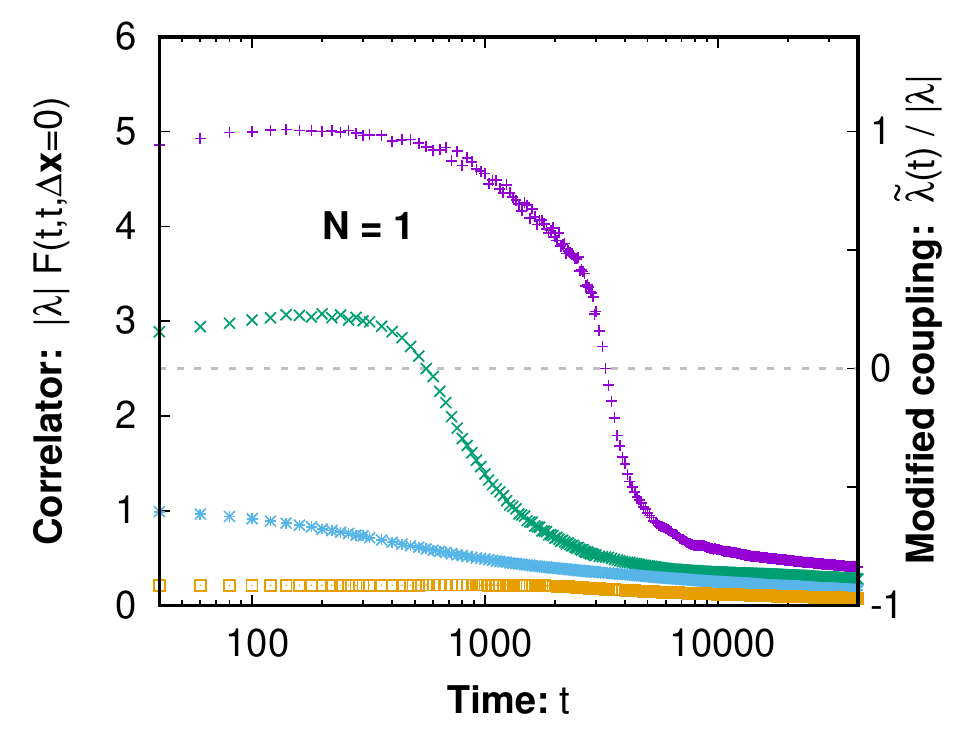}
    \includegraphics[scale=0.835]{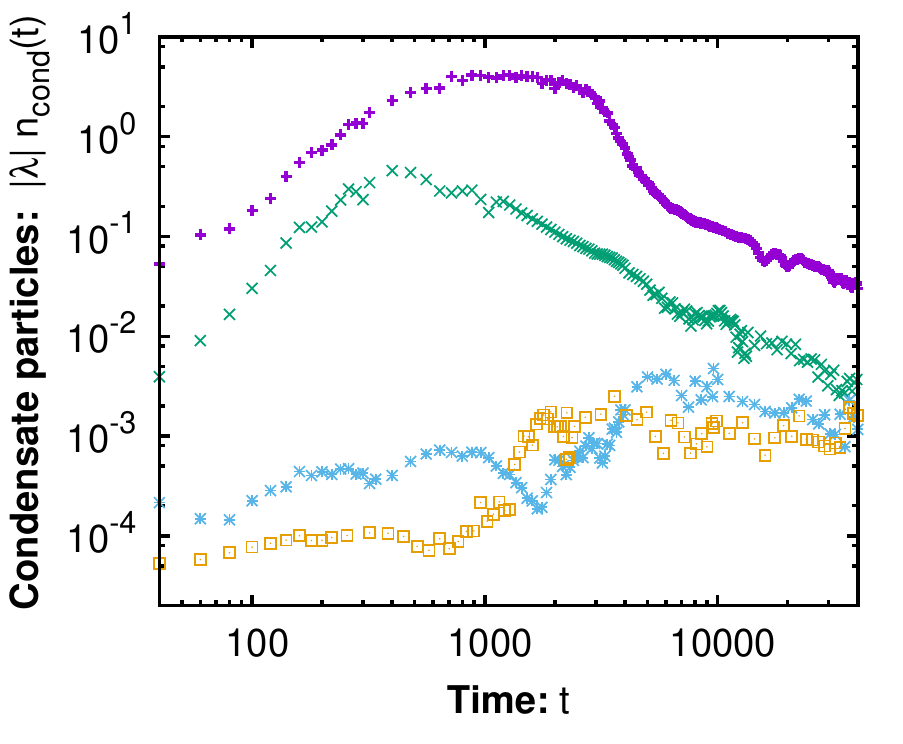}
    \includegraphics[scale=0.835]{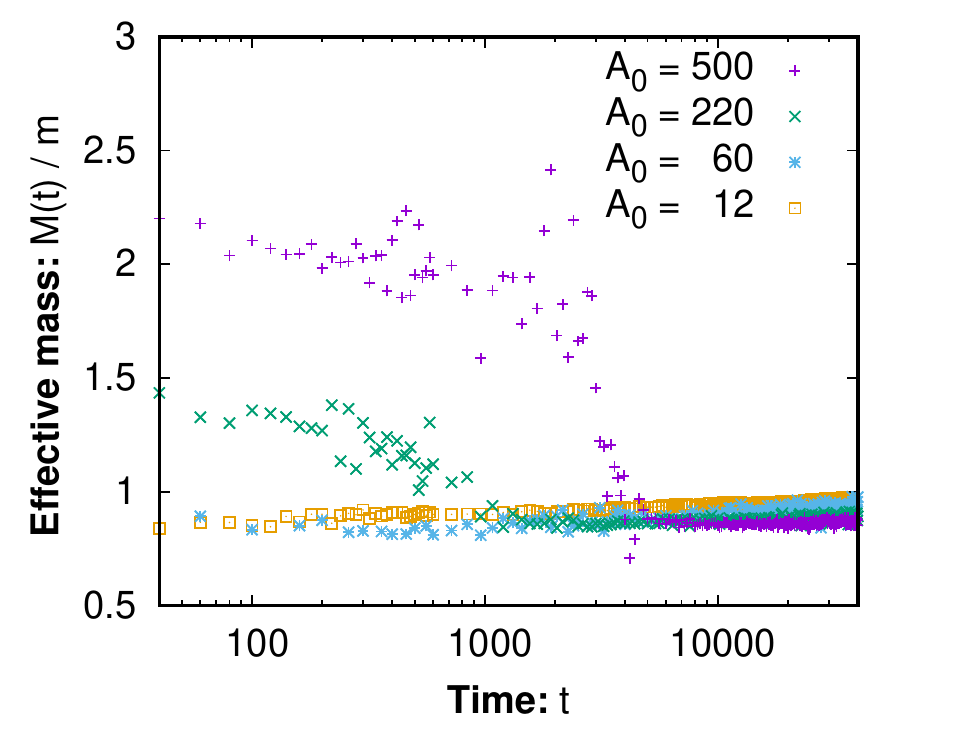}
\caption{{\bf Left: }Total particle number density $n(t)$ ({\bf upper panel}) and particle number density in the condensate $n_{\mathrm{cond}}(t)$ ({\bf lower panel}), defined in Eq. (\ref{eq:tpnd}), as functions of time for several initial occupancies (logarithmic scale). {\bf Right:} The integrated correlator $F(t,t,\mathbf{\Delta x=0})$ ({\bf upper panel}) defined in Eq.~(\ref{eq:FttDx0}) and numerically extracted effective mass $M(t)/m$ ({\bf lower panel}), as functions of time for different initial occupancies. In the upper left panel, the black dashed line indicates the approximate scaling behavior and in the upper right panel the grey dashed line denotes the zero crossing of the modified coupling $\widetilde \lambda(t)$ given by Eq.~(\ref{eq:mod_lambda}).}
\label{fig:transition}
\end{figure}

The decay of the high infrared occupancy and the collapse of the condensate are caused by inelastic processes. Indeed, as can be seen in the upper left panel of Fig.~\ref{fig:transition} for $\occupancy = 220$ and $\occupancy = 500$, the total particle number density decreases with time. This interpretation is further strengthened by the observation of accompanying peaks of the distribution function at momenta $|\mathbf{p}|\gtrsim M(t)$ shown in the lower left panel of Fig.~\ref{fig:rep-attractive}, where $M(t)$ is the effective mass. The locations of these peaks correspond to $2n \rightarrow 2$ inelastic scatterings off the ``condensate'' or off modes with momenta considerably below $M(t)$. Such decay peaks have also been observed in quartic models \cite{Moore:2015adu, Berges:2013lsa}. For instance, the sharp peak of the $t=8000$ curve is located at momentum $|\mathbf{p}| \approx \sqrt{3}M \approx 1$,\footnote{The value of the effective mass at $t=8000$ is $M \approx 0.6$, which can be deduced from the lower right panel of Fig.~\ref{fig:transition}, using the mass parameter $m=0.7$.} which corresponds to an inelastic scattering of four particles with $|\mathbf{p}| \ll M$ into two particles with energy $\epsilon \approx 2M$. The peaks move towards smaller momenta with time. This is a consequence of the particle loss, which leads to the decrease of the effective mass (see lower right panel of Fig.~\ref{fig:transition}). Moreover, the peaks become more pronounced as the height of the initial box is increased, since in this case the condensate and low momentum modes contain more particles that can undergo inelastic scatterings.

Importantly, the particle loss generates a transition from repulsive to attractive mean interactions. An important quantity which characterizes the strength of the fluctuations, and visualizes the transition, is the integrated correlator 
\beq
 \label{eq:FttDx0}
 F(t,t,\mathbf{\Delta x=0}) = \int \frac{d^3p}{(2\pi)^3}\, F(t,t,\mathbf{p})\,.
\eeq
This correlation function characterizes the strength of the fluctuations and determines shifts to the effective mass and the coupling. As can be seen from its definition (\ref{eq:FttDx0}), it is an integral over all momentum modes and is additionally dominated by the large infrared modes, which can be understood from footnote \ref{fn:definition_distrib}. Its decrease is a consequence of particle loss. The time evolution of this quantity is shown in the upper right panel of Fig.~\ref{fig:transition} for different initial occupancies. As can be seen, for large values of $\occupancy$ ($220$ and $500$ in the plot) there is a stage when the correlator decreases rapidly. This stage represents the transition between the two regimes and corresponds to the rapid decrease of the total particle number density, the collapse of the condensate and the effective mass $M(t)$ falling below its bare value $m$, as seen in the residual panels of the figure. Around this time window, the observed decay peaks in the particle spectrum that were discussed above become more pronounced. 

In addition, we show the zero crossing of a modified quartic coupling
\beq
 \label{eq:mod_lambda}
 \widetilde \lambda(t) = \lambda + \frac{\lambda_6}{10}\, F(t,t,\mathbf{\Delta x=0})
\eeq
as a grey dashed line in the upper right panel of Fig.~\ref{fig:transition}. This time-dependent quartic coupling results from a systematic $1/N$ expansion to next-to-leading order and will be derived in Sec.~\ref{sec:2PI_1N_kinetic}. There we will show that to this order, the coupling parameter $\lambda$ should essentially be replaced by $\widetilde \lambda(t)$ in scattering processes of scalar theories with a large $N$. Interestingly, this modified coupling changes its sign during the transition, from positive to negative, as can be seen in the figure. This is a remarkable observation given the fact that the modified coupling is derived in a $1/N$ expansion while we consider a single-component model here. Similar observations will be made in Sec.~\ref{sec:class_stat} for scalar theories with $N \geq 2$ field components. 

$\bigskip$ 

In summary, although the dynamics of the infrared fixed point is insensitive to the precise form of the self-interactions, it does depend on whether mean self-interactions are repulsive or attractive. While repulsive self-interactions have the effect of moving the system closer to the infrared nonthermal fixed point, which also leads to condensation, the presence of attractive self-interactions has the opposite effect. Such interactions are accompanied by enhanced annihilation processes, which prevent the formation of large structures. In contrast, an energy cascade to higher momenta occurs in both cases and is, in this sense, a more generic phenomenon.


\section{Dynamics for multi-component fields ($N>1$)}
\label{sec:dynamics_N_comp}

In this section, we study the dynamics of our model in the case when the number of components is larger than one. Our analysis includes two different nonperturbative methods. One of them is classical-statistical simulations, similar to those that were employed in the previous section. Many aspects of the dynamics turn out to be universal and do not indicate strong dependence on $N$. In addition to this, we develop a vertex-resummed kinetic theory for our model in Section~\ref{sec:2PI_1N_kinetic}, based on a $1/N$ expansion of the 2PI quantum effective action up to the next-to-leading order (NLO). The $1/N$ expansion relies on a large number of field components and is based on a classification of the contributions to the effective action according to their scaling with $N$. This provides a controlled expansion parameter, which is not restricted to weakly coupled or low occupied systems. The 2PI $1/N$ expanded theory to NLO and the vertex-resummed kinetic theory have been successfully applied to repulsive quartic $\mathcal{O}(N)$-symmetric scalar field theories and were able to describe observed phenomena \cite{Berges:2001fi, Aarts:2002dj, Berges:2015kfa}. Therefore, these approximations also seem to be a suitable approach to the case of attractive interactions.

\subsection{Classical-statistical simulations}
\label{sec:class_stat}

We extend here the analysis of Section~\ref{sec:dynamics_N1} to the $N>1$ case. More specifically, we consider two-, four- and eight-component fields and use lattices with up to $512^3$ points and lattice spacings in the range of $0.1$ - $0.5$ $\mathrm{p_0^{-1}}$. We checked that shown results are insensitive to the lattice spacing and grid size. The details of the numerical calculations can be found in Appendix A. We employ $g^2 = 2$ and $m = 0.7\, \mathrm{p}_0$, with $\mathrm{p}_0$ introduced by the initial conditions (\ref{eq:box}), and rescale all dimensionful quantities in the figures with appropriate powers of $\mathrm{p}_0$. The repulsive and attractive models are discussed separately.

\subsubsection{Repulsive ($\lambda>0$)}
\label{sec:N_comp-class_stat-rep}

As in the quartic theory \cite{Orioli:2015dxa}, the basic features of the time evolution of the momentum distribution function $f(t,|\mathbf{p}|)$ reveal weak dependence on the number of field components. Similar to the $N=1$ case, two cascades emerge after some transient time: an inverse particle cascade at low momenta and a direct energy cascade at high momenta. The inverse cascade becomes self-similar, with the same scaling exponents $\alpha\approx 3/2$ and $\beta\approx 1/2$ and the same scaling function $f_S(p)$ that were observed in the single-component model, and leads to the formation of a condensate. The dynamics at low momenta is dominated by elastic scatterings, so that the total particle number (\ref{eq:tpnd}) is approximately conserved.

There are some differences in the spatial structure of the condensate for different $N$, which, however, do not affect the above mentioned universal self-similar dynamics. They can be studied directly by looking at the classical field $\varphi_a(t, \mathbf{x})$ in position space of a single simulation, before taking the ensemble average. This has been done in \cite{Moore:2015adu} for the repulsive $\varphi^4$ theory. In the single-component model, the condensate in one simulation represents a coherently ``back-and-forth'' oscillating classical field. For $N \geq 2$, more general orbital forms in field component space for coherent oscillations are possible. Circular orbits turn out to be energetically preferred \cite{Moore:2015adu} and, as a result, in multi-component models the coherent oscillations are circular. 

\begin{figure}[tp!]
\centering
\includegraphics[scale=0.47]{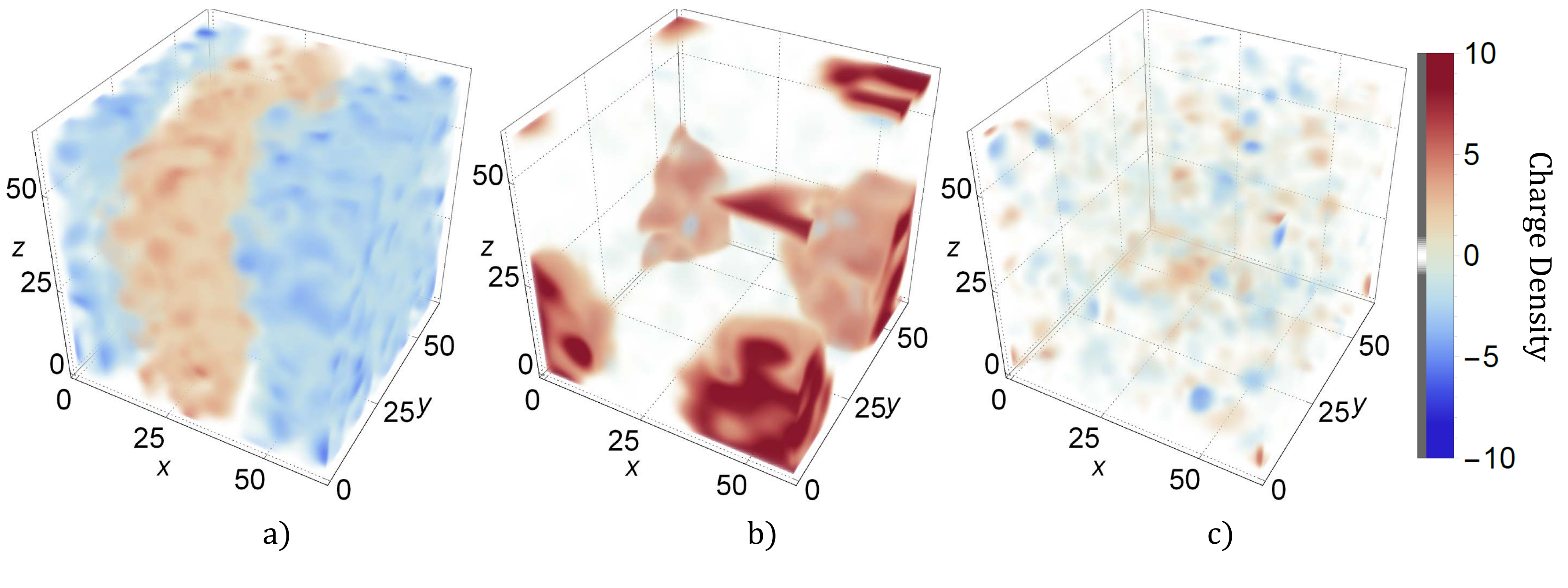}
\caption{Snapshots of the charge density $\rho(\mathbf{x})=\varphi_1 (\mathbf{x})\dot{\varphi}_2(\mathbf{x}) - \varphi_2(\mathbf{x}) \dot{\varphi}_1(\mathbf{x})$ at $t=5000$ for two-component field theory with $\occupancy=60$ and (a) $\lambda>0$ without charge asymmetry, (b) $\lambda<0$  with charge asymmetry and (c) $\lambda<0$ without charge asymmetry.}
\label{fig:charge}
\end{figure}

This may be further understood by the presence of continuous $\C{O}(N)$-symmetries for $N \geq 2$ that impose additional restrictions on the field \cite{Moore:2015adu}. Importantly, there are $N(N-1)/2$ conserved Noether charges of the form
\beq
\label{eq:Noether}
Q_{ab}=\int d^3 \mathbf x \left(\varphi_a \dot{\varphi}_b - \varphi_b \dot{\varphi}_a \right) \qquad \qquad (a>b),
\eeq
associated with global rotations in the corresponding planes in field space. For our considered initial conditions, each of these charges vanishes (see Appendix A). However, the charges may differ from zero locally. As a result, for $N=2$ the condensate splits into different domains, each of which has a positive or negative charge (see Fig.~\ref{fig:charge}a) depending on the direction of the circular oscillations, such that the total charge vanishes. These domains are separated from each other by thin walls, also visible in the figure, where the field oscillates linearly and carries no charge. While number changing processes within each domain are suppressed because of charge conservation, the presence of domain walls leads to a small leakage of particle number.\footnote{Similar to the single-component case, a nonzero bare mass further suppresses particle number changing processes \cite{Moore:2015adu}.} For $N>2$ there is a continuous group of possible oscillations and domain walls are absent there. Therefore, annihilation processes are even stronger suppressed. We refer to Ref.\cite{Moore:2015adu} for a detailed discussion of topological defects for different $N$.  

\subsubsection{Attractive mean interactions with $\lambda<0$}
\label{sec:mean_attr}

When the quartic coupling is negative, two qualitatively different regimes of the dynamics can be observed, depending on which of the two self-interactions is more important. For relatively weak fluctuations mean interactions are attractive, while for strong fluctuations they are repulsive. The dynamics in both regimes proceeds along the same lines as for $N=1$.

In the case of mean attraction, the inverse cascade, which was observed in the repulsive model, is absent and no condensate is formed. The low momentum modes grow for some transient time and decay slowly afterwards. Also the total particle number density soon decreases approximately as a power-law $\sim t^{\beta'}$ with $\beta' \approx -1/5$, being the same as discussed in Sec.~\ref{sec:dynamics_N1} for $N=1$. Hence, number changing processes are important also for $N > 1$ with mean attractive interactions.

Although in single simulations small clumps, that typically contain positive and negative charge, can develop during the evolution, they quickly decay via particle annihilation and thus, are not stable. A typical late-time configuration is shown in Fig.~\ref{fig:charge}c for $\occupancy=60$ at $t=5000$. This is in notable difference to the non-relativistic case, which features a conserved charge (particle number) and accordingly leads to the formation of persistent clumps (see, e.g.~\cite{Khlebnikov:1999qy})\footnote{For the case of axions long range gravitational interactions are supposed to be the dominant attractive interaction. In this case it is not a priori clear if the number changing processes are similarly effective, or if there exists an approximate particle number conservation.}.
\bigskip

As already mentioned, our considered initial conditions generate no $\C{O}(2)$ charge. However, with a slight modification of the initial conditions, which on the level of a single simulation consists in correlating the initial oscillation phases of each of the field components, a non-vanishing charge $Q$ can be generated. This is explained in Appendix A, where it is also shown that the value of the charge can be controlled by the height $A_0$ of the initial distribution. The charge density of a typical late-time configuration with these modified initial conditions is shown in Fig.~\ref{fig:charge}b for the case of $\occupancy=60$ at time $t=5000$. As can be seen, most of the system's charge is concentrated in an almost spherical object, which corresponds to a so-called $Q$-ball~\cite{Coleman:1985ki}.\footnote{After some time several smaller Q-balls add up to a single large Q-ball, that is visible in the figure.}
This is in stark contrast to our usual choice of initial conditions (\ref{eq:box}) without an initial charge and more similar to the non-relativistic model with a conserved particle number.

Indeed, it is a well-known feature of potentials, which contain an attractive self-interaction and thus grow slower as compared to quadratic in some range of field values, that the presence of a Noether charge associated with an unbroken global symmetry leads to the existence of compact lowest-energy field configurations, i.e.~$Q$-balls \cite{Coleman:1985ki}. They are spherically symmetric in coordinate space and at each spatial point the classical field oscillates along circular orbits in field space. $Q$-balls are non-topological solitons since their stability against decay is guaranteed by the conserved charge they carry. For the simplest case of $N=2$, a $Q$-ball has the form (in polar coordinates with an appropriate choice of the origin) 
\begin{equation}
\label {eq:Qballsphsym}
\varphi_1 = \phi(r) \cos(\omega t),\qquad \varphi_2 = \phi(r) \sin(\omega t),
\end{equation}
where $\phi(r)$ determines the radial profile. $Q$-balls with a large charge can be described using the thin-wall approximation \cite{Coleman:1985ki} and their profile function has a form close to a step function: $\phi(r)=\phi_Q \Theta(R-r)$, with a constant amplitude $\phi_Q$ and with $R$ denoting the radius. In contrast, $Q$-balls with small charge have thick walls, and there is evidence that they become unstable for a sufficiently small charge \cite{Multamaki:1999an}.

\subsubsection{Repulsive mean interactions with $\lambda<0$}

For sufficiently high initial occupancies mean interactions become initially repulsive. The dynamics in this regime is very similar to the $\lambda>0$ case. In particular, an inverse particle cascade with the same values of the scaling exponents $\alpha \approx 3/2$ and $\beta \approx 1/2$ is again observed and leads to the formation of a condensate. We have seen in Sec.~\ref{sec:N1_attractive} that in the single-component model, particle number was not conserved and the particle loss gradually increased the role of the (attractive) quartic self-interaction. This eventually generated a transition to the mean attractive regime, accompanied by a fragmentation of the condensate. We observe a similar behavior for $N=2$. However, the annihilation rate is suppressed as compared to $N=1$ and, as a result, the lifetime of the condensate is longer. The structure of the condensate plays an important role for this suppression \cite{Moore:2015adu}. As discussed in Sec.~\ref{sec:N_comp-class_stat-rep}, for $N=2$ the classical field locally carries maximal charge, which means that particle annihilation from the same domain of the condensate would violate charge conservation. The domain walls, that are necessarily present in this case, are therefore expected to be the dominant source of particle evaporation. For larger values of $N$, such as $N=4$ or $N=8$, number changing processes in the repulsive regime are even stronger suppressed, which makes the condensate long-lived. 

\begin{figure}[tp!]
  \centering
    \includegraphics[scale=1.214]{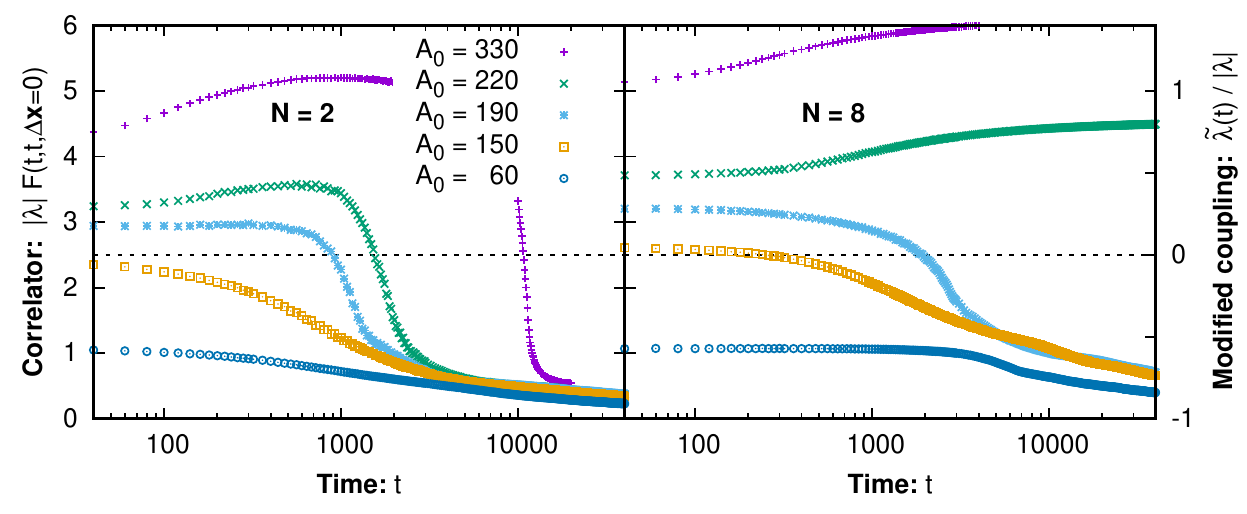}
\caption{The integrated correlation function $F(t,t,\mathbf{\Delta x=0})$ defined by Eq.~(\ref{eq:FttDx0}) as a function of time for several initial occupancies for $N=2$ (left) and $N=8$ (right). The dashed line in each panel marks the zero crossing of the modified quartic coupling $\widetilde \lambda(t)$ of Eq.~(\ref{eq:mod_lambda}) that is derived in Sec.~\ref{sec:2PI_1N_kinetic} and shown in the right y axis of the right panel.}
\label{fig:transition28}
\end{figure}

To demonstrate the transition, we plot the integrated correlation function $F(t,t,\mathbf{\Delta x=0})$ (\ref{eq:FttDx0}) as a function of time in Fig.~\ref{fig:transition28} for several values of the initial occupancy $A_0$. The two panels correspond to $N=2$ (left) and $N=8$ (right). As in the $N=1$ case (see Fig.~\ref{fig:transition}), the region where $F(t,t,\mathbf{\Delta x=0})$ decreases rapidly corresponds to the collapse of the condensate and approximately separates the mean repulsive from the mean attractive regime. This is also demonstrated by a black dashed line that shows the zero crossing of the modified coupling $\widetilde \lambda(t)$ of Eq.~(\ref{eq:mod_lambda}). The modified coupling indeed changes its sign during the fast transition, similarly to what was observed for the $N=1$ case. Since the derivation of $\widetilde \lambda(t)$ is based on a $1/N$ expansion, as will be shown in the following section, the modified coupling is expected to provide a good description for theories with many field components. That its zero crossing mostly falls into the transition region also for theories with few field components, is an unanticipated but important observation that supports the interpretation in terms of repulsive and attractive mean interactions. 

In addition, one observes another interesting phenomenon in Fig.~\ref{fig:transition28}. For $N=2$ the transition occurs for any of the considered amplitudes of the initial occupancy within the simulation time, occurring later for higher amplitudes. Compared to $N=2$ for $N=8$ it generally occurs at later times. Remarkably there seems to be a rather dramatic increase between $\occupancy=190$ (blue) and $\occupancy=220$ (green), such that the transition does not occur within the simulation time. The system remains in the mean repulsive regime and the modified coupling stays positive, $\widetilde \lambda(t) > 0$. This difference between $N=2$ and $N>2$ theories is consistent with the argument about enhanced suppression of annihilation processes for $N>2$ as compared to $N=2$ theories in the end of Sec.~\ref{sec:N_comp-class_stat-rep}. 

Finally, if the initial conditions are modified to generate a non-vanishing charge density, as explained in Sec.~\ref{sec:mean_attr} and Appendix \ref{app:class_stat}, a condensate is formed for sufficiently high initial occupancies. It consists of a single charge domain and number changing processes are forbidden. As a result, the condensate is stable and no transition to the mean attractive regime occurs in this case. This is consistent with the picture that for very high charge densities the condensate corresponds to a circular orbit in field space with an amplitude such that the field is always in the repulsive regime according to Fig.~\ref{fig:potential}.


\subsection{Vertex-resummed kinetic theory from 2PI $1/N$ expansion}
\label{sec:2PI_1N_kinetic}

To understand the above observations from classical-statistical simulations, we derive in the following an effective kinetic description that is based on a $1/N$ expansion to next-to-leading order (NLO). As it turns out, the description is similar to the vertex-resummed kinetic theory that has been used to describe scaling phenomena in repulsive quartic scalar theories \cite{Berges:2010ez, Orioli:2015dxa}. The main difference is the emergence of a modified coupling $\widetilde \lambda(t)$, which incorporates corrections from the presence of the sextic coupling. As mentioned in previous sections, the modified coupling varies with time and may even change its sign, which is then accompanied by a transition from repulsive to attractive mean interactions.


\subsubsection{2PI $1/N$ expansion to NLO}
\label{sec:2PI_1N_exp}

We start with the full quantum field theory whose properties are encoded in the two-particle-irreducible (2PI) effective action $\Gamma[\phi, G]$ \cite{Cornwall:1974vz}. It is a free energy functional, parametrized in terms of the macroscopic field $\phi_a(x)$ and the time-ordered connected two-point function $G_{ab}(x,y)$, also referred to as the (full) propagator and given by
\beq
\label{eq:definition_G}
G_{ab}(x,y) = \langle \mathrm{T}_C \hat \varphi_a(x) \hat \varphi_b(y) \rangle - \phi_a(x)\phi_b(y).
\eeq
The time variables are defined on the Schwinger-Keldysh contour $C$ \cite{Keldysh:1964ud}, which starts at $t=0$, runs forwards along the real-time axis and then returns back to $t=0$. Here $\mathrm{T}_C$ denotes the time-ordering operator along the time contour $C$. The correlation functions $\phi$ and $G$ satisfy the quantum equations of motion that are given by stationarity conditions of the effective action,
\beq
\frac{\delta \Gamma }{\delta \phi_a(x)}=0, \qquad
\frac{\delta \Gamma }{\delta G_{ab}(x,y)}=0.
\label{statconditions}
\eeq
The full propagator (\ref{eq:definition_G}) is a complex-valued function and it is convenient to separate its real and imaginary parts. They can be associated with the expectation values of the commutator and anti-commutator of the field operators. The decomposition is given by \cite{Aarts:2001qa}
\beq
G_{ab}(x,y) \: = \: F_{ab}(x,y) \: - \: \frac{i}{2}\rho_{ab}(x,y)\,
\textrm{sgn}_{C}(x^{0}-y^{0}),
\label{separation}
\eeq
where
\begin{eqnarray}
  F_{ab}(x,y) &=& \frac{1}{2} \langle \{ \hat{\varphi}_{a}(x),  \hat{\varphi}_{b}(y) \}\rangle  - \phi_a(x)\phi_b(y), \\
 \rho_{ab}(x,y) &=&  i \langle [ \hat{\varphi}_{a}(x),  \hat{\varphi}_{b}(y) ] \rangle,
\label{separationParts}
\end{eqnarray}
and $\mathrm{sgn}_C(x^{0}-y^{0})$ is $\pm1$ depending on whether $x^0$ is after or before $y^0$ along the closed time contour. $F$ is the statistical propagator and $\rho$ denotes the spectral function. Both are real-valued functions with different symmetry properties: $ F_{ab}(x,y)= F_{ba}(y,x)$, $\rho_{ab}(x,y)=-\rho_{ba}(y,x)$. 

The contributions to the 2PI effective action can be written as \cite{Cornwall:1974vz}
\beq
  \Gamma [\phi,  G]  =    S_C[\phi]  +  \frac{i}{2} \mathrm{Tr}_{C} \ln G^{-1} +  \frac{i}{2} \mathrm{Tr}_{C} \{ G_{0}^{-1}(\phi)G \} + \Gamma_{2}[\phi,  G]  +  \mathrm{const}.
\label{eq:expression_for_effaction}
\eeq
Here the traces include spacetime integration\footnote{The subscript $C$ corresponds to an integration of the time variable along the Schwinger-Keldysh contour $C$.} as well as summation over field indices, e.g. $\mathrm{Tr}_{C} \{ G_{0}^{-1}(\phi)G \}= \int_{xy, C} G_{0\: ab}^{-1}(x,y; \phi) G_{ba}(y,x)$. The function $G_{0}^{-1}$ is the inverse classical propagator, defined as 
\beq
\label{eq:clasprop}
iG_{0\: ab}^{-1}(x,y;\phi)=\frac{\delta^2S_C[\phi]}{\delta\phi_a(x) \delta \phi_b(y)}.
\eeq
The 2PI functional $\Gamma_2[\phi,G]$ includes all diagrams, that do not become disconnected after removing two inner (full) propagator lines. This functional is the sum of all such 2PI diagrams with full propagators $G$ and vertices given by the non-linear and non-quadratic (with respect to $\varphi$) parts of $S_C[\phi+\varphi]$, which we denote by $S_{\mathrm{int}}[\varphi; \phi]$. 

Setting $\phi=0$, the inverse classical propagator and the interaction term for our model (\ref{eq:potential}) are given by
\begin{eqnarray}
  \label{eq:propagator_cl}
 iG^{-1}_{0\: ab} (x,y)   &=&   -  \Bigl( \Box_{x}  +m^{2} \Bigr)\delta_{ab}  \delta (x-y),  \\
 S_{\mathrm{int}}[\varphi]  &=&   - \int_{x, C}\Biggl[\frac{\lambda}{4!N} \Bigl( \varphi_{b}(x) \varphi_{b}(x) \Bigr)^{2} + \frac{\lambda_6}{6!N^{2}} \Bigl( \varphi_{b}(x) \varphi_{b}(x) \Bigr)^{3}   \Biggr].
\end{eqnarray}
In a practical computation, to solve the equations of motion (\ref{statconditions}) the effective action needs to be truncated. This can be done in a nonperturbative way using a $1/N$ expansion, by classifying the contributions to $\Gamma_2[\phi = 0,G]$ based on their scaling with the number of field components $N$. The diagrams up to NLO have been derived in Refs.~\cite{Townsend:1975kh,Townsend:1976sy}. 

\begin{figure}[t]
\center{\includegraphics[scale=0.47]{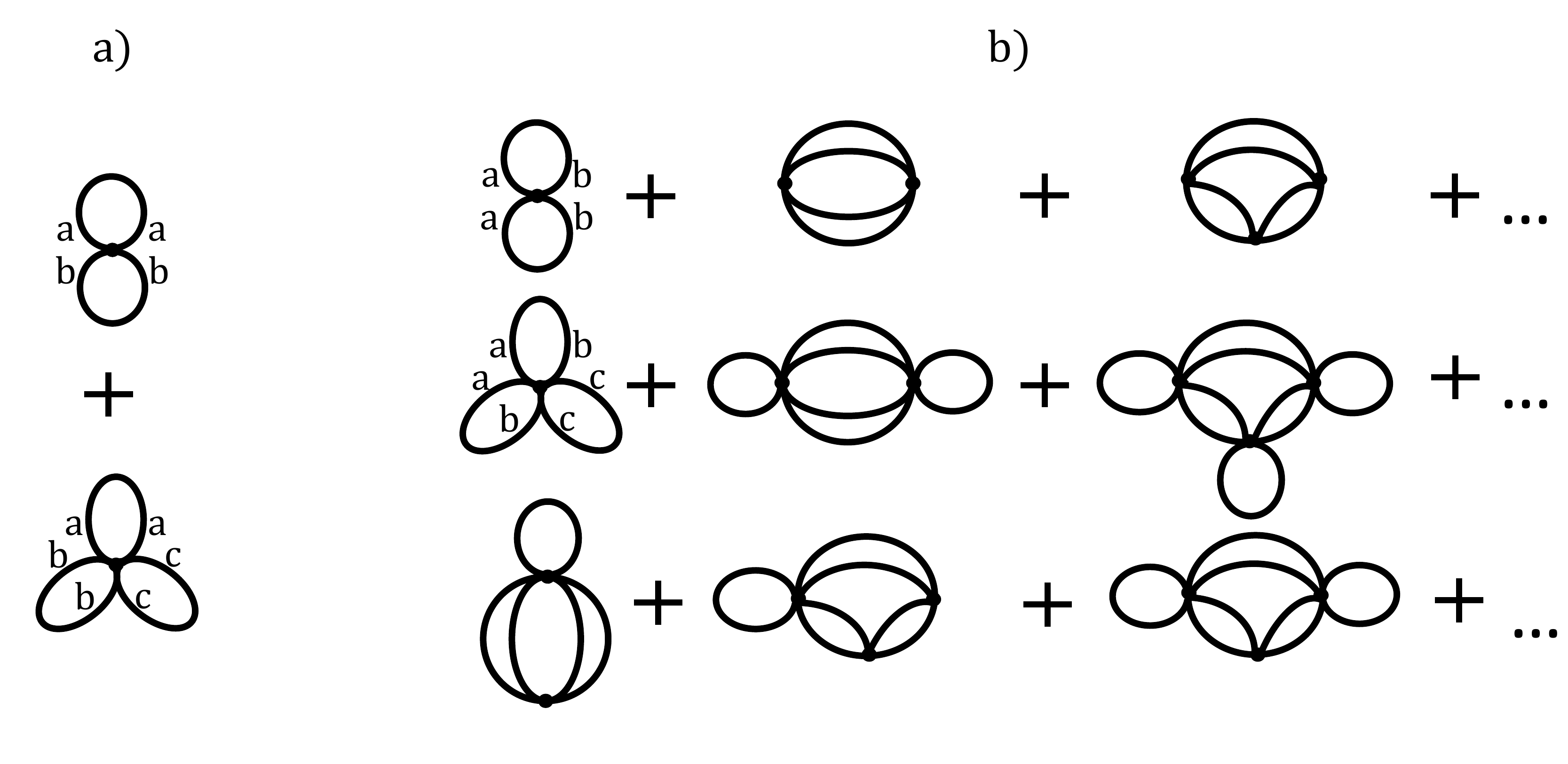}}
\caption {Leading order (a) and next-to-leading order (b) diagrams}
\label{fig:diagrams}
\end{figure}

There are only two vacuum graphs at leading order~\cite{Townsend:1975kh}, shown in Fig.~\ref{fig:diagrams}a. These correspond to diagrams with only one vertex where propagators $G_{aa}(x,x)$ connect legs with the same indices forming closed bubbles. Because of the implicit summation over the index $a$, each bubble, and consequently each of the two shown diagrams, scales proportional to $N$. Therefore, we obtain
\beq
 \Gamma_{2}^{\mathrm{LO}}[G] \: = \: - \: \frac{\lambda}{4!N} \int_{x,C} G_{aa}(x,x)G_{bb}(x,x) \:
 - \: \frac{\lambda_6}{6!N^{2}} \int_{x,C} G_{aa}(x,x) G_{bb}(x,x) G_{cc}(x,x).
\label{eq:GammaLO}
\eeq
Both terms in (\ref{eq:GammaLO}) contribute only as a coordinate-dependent shift to the effective mass and, in particular, at LO no scattering processes are included (see Appendix B). The NLO contribution consists of an infinite series of diagrams \cite{Townsend:1976sy}, shown in Fig.~\ref{fig:diagrams}b. The first row contains all diagrams without sextic vertices. It is the complete NLO contribution in the $\varphi^4$ theory and has been resummed in \cite{Berges:2001fi}. In the two lower rows, diagrams include at least one sextic vertex. The latter, whenever encountered, necessarily contains one closed bubble $G_{aa}(x,x)$ to compensate its additional factor of $1/N$, while the rest of its legs form the same propagator chains as in the first row. Note that the two diagrams appearing at LO are also encountered at NLO, however, with a different index structure, as explicitly shown in Fig.~\ref{fig:diagrams}.

To combine the diagrams of the first row with those containing sextic vertices, it is convenient to introduce a coordinate-dependent ``modified'' coupling\footnote{In Eq. (\ref{eq:lmod}) we have used the fact that $\rho_{aa}(x,x)=0$ and we have defined $F(x,x) = F_{aa}(x,x)/N$.}
\beq
\widetilde \lambda(x)= \lambda \: + \: \frac{\lambda_6}{10N}G_{aa}(x,x)= \lambda \: + \: \frac{\lambda_6}{10}F(x,x),
\label{eq:lmod}
\eeq
which plays the role of the quartic coupling at NLO in presence of the sextic self-interaction. With this, the resummation of the NLO diagrams proceeds in the same way as for the $\varphi^4$ theory and yields \cite{Townsend:1976sy}
\beq
\label{eq:GammaNLO}
\Gamma_{2}^{\mathrm{NLO}}[G]\: = \: {\frac{i}{2}}\: \textrm{Tr}_{C} \ln[\widetilde B(G)],
\eeq
where we have defined $\widetilde B(x,y;G)=  \delta(x-y)+\: i\frac{\widetilde \lambda(x)}{6N}G^2(x,y)$, with $G^2(x,y) = G_{ab}(x,y)G_{ab}(x,y)$, and the logarithm sums the infinite series
\begin{equation}
\label{eq:gamma}
  \textrm{Tr}_{C} \ln[\widetilde B(G)] = {\int_{x,C}} \: i \: {\frac{\widetilde \lambda(x)}{6N}} \: G^2(x,x) \: -  \: {\frac{1}{2}} \: {\int_{xy,C} } \: \Bigl( \: i \: {\frac {\widetilde \lambda(x)}{6N}} \: G^2(x,y) \Bigr) \: \cdot \: \Bigl( \: i \: {\frac {\widetilde \lambda(y)}{6N}} \:  G^2(y,x) \Bigr) + \dots
\end{equation}

The arguments of this section can be generalized to any higher order self-interactions of even power. For a general interaction $S_{\mathrm{int}}[\varphi]=-\int \sum_{n=2}^{\infty}\frac{\lambda_{2n}}{(2n)!N^{n-1}} \Bigl(\varphi_a(x)\varphi_a(x)\Bigr)^{n}$, the NLO modified quartic coupling is given by
\beq
\widetilde \lambda(x) = \sum_{n=2}^{\infty} \frac{6(n-1)\lambda_{2n}}{(2n-1)! } \Bigl( F(x,x) \Bigr) ^ {n-2}.
\label{eq:lmod2n}
\eeq

Stated differently, at NLO the higher power self-interactions contribute as a coordinate-dependent shift to the quartic coupling, given by (\ref{eq:lmod}) and (\ref{eq:lmod2n}), while topologically the diagrams are the same as for the quartic scalar theory. In general, the modified coupling can be positive or negative, depending on the signs of the coupling parameters and on the magnitude of $F(x,x)$, which is connected to the system's occupation numbers. We have already shown the time evolution of the modified coupling in previous sections in Figs.~\ref{fig:transition} and \ref{fig:transition28} for the sextic theory with attractive quartic interaction and we have observed qualitatively different dynamics depending on the sign of the modified coupling. To get closer to an interpretation in terms of quasiparticles, we derive the corresponding kinetic theory in the following.


\subsubsection{Effective kinetic equation}

In a kinetic framework, the time evolution of the distribution function is given by the Boltzmann equation. It has the form
\beq
\frac{\partial f(t,\mathbf{p})}{\partial t} = C[f](t,\mathbf{p}),
\eeq
where the collision integral $C[f]$ represents a sum over all possible scatterings with in- or outgoing momentum $\mathbf{p}$. For instance, in a coupling expansion, the leading order perturbative expression for the $2 \leftrightarrow 2$ collision integral in $\varphi^4$ theory is given by \cite{Berges:2015kfa}
\beq
C^{2 \leftrightarrow 2} [f] (t,\mathbf{p}) = \int d\Omega^{2\leftrightarrow 2}[f](t,\mathbf{p},\mathbf{q},\mathbf{r},\mathbf{l}) [(f_{\mathbf{p}}+1)(f_{\mathbf{l}}+1)f_{\mathbf{q}}f_{\mathbf{r}}-f_{\mathbf{p}}f_{\mathbf{l}}(f_{\mathbf{q}}+1)(f_{\mathbf{r}}+1)],
\eeq
with the shortcut notation $f_{\mathbf{p}} = f(t,{\mathbf{p}})$, the Bose enhancement factors $f_{\mathbf{p}}+1$, and, writing $\int_{\mathbf{q}} = \int d^3 \mathbf{q}/(2\pi)^3$, with the integration measure
\beq
\int d\Omega^{2\leftrightarrow 2}[f](t,\mathbf{p},\mathbf{q},\mathbf{r},\mathbf{l})=\lambda^2\frac{N+2}{6N^2} \int_{\mathbf{lqr}}(2\pi)^{4} \delta(\mathbf{p}+\mathbf{l}-\mathbf{q}-\mathbf{r}) \frac{ \delta (\omega_\mathbf{p}+\omega_\mathbf{l}-\omega_\mathbf{q}-\omega_\mathbf{r}) }{ 2\omega_\mathbf{p} 2\omega_\mathbf{l} 2\omega_\mathbf{q} 2\omega_\mathbf{r} }.
\label{eq:perturbbolotz}
\eeq

\begin{figure}[!t]
  \centering
    \includegraphics[scale=0.45]{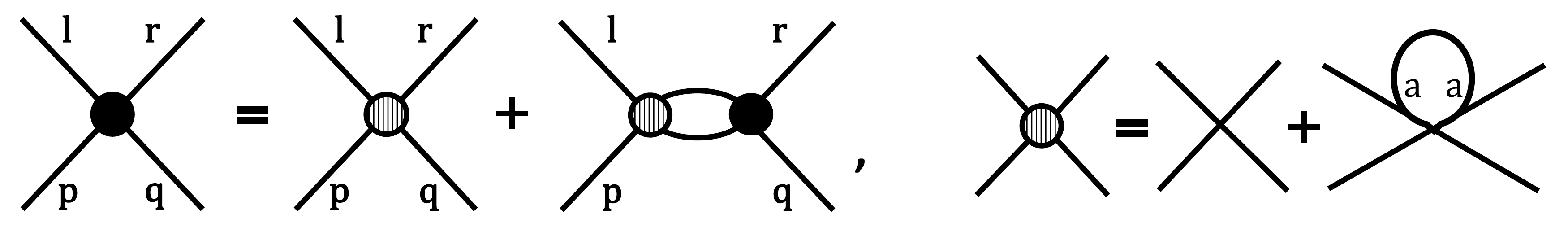}
\caption{On the left hand side, the self-consistent equation for the s-channel part of the effective coupling is depicted. Filled circles represent the NLO effective four-vertex (\ref{eq:leff}) while dashed circles correspond to the modified coupling $\widetilde \lambda (x)$ from (\ref{eq:lmod}), whose diagrammatic representation is shown on the right hand side.}
\label{fig:effectivevertex}
\end{figure}

Such a perturbative approach is not expected to provide an accurate description of the time evolution of the highly occupied infrared modes, which dominate the dynamics in our model. Because of $|\lambda| f \gtrsim 1$, infinitely many $2 \leftrightarrow 2$ scattering diagrams become as important as the perturbative expression (\ref{eq:perturbbolotz}). Remarkably, kinetic theory can still be extended to the highly occupied regime, with the help of a systematic resummation of scattering processes by use of the $1/N$ expansion to NLO that was discussed above. This has been done for the $\varphi^4$ theory \cite{Berges:2015kfa, Berges:2005md, Berges:2010ez} by deriving the corresponding truncated 2PI equations of motion and recasting them into a form similar to the Boltzmann equation, under additional assumptions such as a comparably smooth time evolution, an effective memory loss and on-shell quasiparticles. In this section, we generalize that derivation to our case. Since at NLO our model can be viewed as a $\varphi^4$ theory with a coordinate-dependent (more specifically time-dependent) quartic coupling, the form of the Boltzmann equation is similar to that of the $\varphi^4$ theory \cite{Berges:2015kfa}, with the replacement $\lambda \mapsto \widetilde \lambda$. The details of the derivation are presented in Appendix \ref{app:transport_equ}. Here we state the final expression for the integration measure of the Boltzmann equation,\footnote{Note that it includes only particle number conserving $2 \leftrightarrow 2$ scatterings. The first on-shell number changing processes, such as $4 \rightarrow 2$ scatterings, appear only at NNLO in the $1/N$ expansion. Off-shell inelastic processes, such as $3 \rightarrow 1$ processes, exist in the 2PI equations of motion already at NLO \cite{Berges:2010ez}, where they are found to be important for final thermalization \cite{Tsutsui:2017uzd}. However, they are not taken into account in the Boltzmann equation.}
\beq
\label{eq:kernel_resummed}
\int d\Omega^{2\leftrightarrow 2}[f](t,\mathbf{p},\mathbf{q},\mathbf{r},\mathbf{l})=\int_{\mathbf{lqr}}\frac{\widetilde \lambda_{\mathrm{eff}}^2(t,\mathbf{p},\mathbf{q},\mathbf{r},\mathbf{l})}{6N} (2\pi)^{4} \delta(\mathbf{p}+\mathbf{l}-\mathbf{q}-\mathbf{r}) \frac{ \delta (\omega_\mathbf{p}+\omega_\mathbf{l}-\omega_\mathbf{q}-\omega_\mathbf{r}) }{ 2\omega_\mathbf{p} 2\omega_\mathbf{l} 2\omega_\mathbf{q} 2\omega_\mathbf{r} }.
\eeq
A time- and momentum-dependent effective four-vertex has been introduced 
\begin{align}
\widetilde \lambda_{\mathrm{eff}}^2(t,\mathbf{p},\mathbf{q},\mathbf{r},\mathbf{l}) = \frac{\widetilde \lambda^2(t)}{3}\Biggl[ & \frac{1}{|1+\Pi_R(t,\omega_\mathbf{p} + \omega_\mathbf{l},\mathbf{p+l})|^2}\; + \;\frac{1}{|1+\Pi_R(t,\omega_\mathbf{p} - \omega_\mathbf{q},\mathbf{p-q})|^2} \nonumber \\
\;+\;&\frac{1}{|1+\Pi_R(t,\omega_\mathbf{p} - \omega_\mathbf{r},\mathbf{p-r})|^2}\Biggr],
\label{eq:leff}
\end{align}
that is a consequence of the infinite series of 2PI diagrams at NLO (\ref{eq:gamma}). The three terms correspond to different scattering channels, where the $s$ scattering channel ($\mathbf{p+l}$ term) is illustrated in Fig.~\ref{fig:effectivevertex} as a self-consistent diagrammatic equation. Expanding the left equation in the figure, the effective vertex appears to be a geometric series of ``chain'' diagrams with double propagator lines that connect modified couplings $\tilde \lambda(t)$ of Eq.~(\ref{eq:lmod}). Therefore, the expression for the effective vertex in (\ref{eq:leff}) can be regarded as an analytic continuation of the geometric series. The ``one-loop'' retarded self-energy $\Pi_R$ in (\ref{eq:leff}) corresponds to two propagator lines and one modified coupling in this picture, i.e.~a ``chain link'', and can be written as \cite{Orioli:2015dxa}
\begin{align}
\label{eq:onshellPiR}
\Pi_R(t,\omega, \mathbf{p})=\lim _{\epsilon \rightarrow 0}\frac{\tilde\lambda(t)}{12}\int_{\mathbf{q}}\frac{f(t,\mathbf{p-q})}{\omega_{\mathbf{q}}\omega_{\mathbf{p-q}}}
 \Biggl[ & \frac{1}{\omega_{\mathbf{q}}+\omega_{\mathbf{p-q}}-\omega- i \epsilon}+\frac{1}{\omega_{\mathbf{q}}-\omega_{\mathbf{p-q}}-\omega- i \epsilon} \nonumber \\
 +\;& \frac{1}{\omega_{\mathbf{q}}+\omega_{\mathbf{p-q}}+\omega+ i \epsilon}+\frac{1}{\omega_{\mathbf{q}}-\omega_{\mathbf{p-q}}+\omega+ i \epsilon} \Biggr].
\end{align}

Having found the corresponding Boltzmann equation, we now discuss some of its properties. In the case that the quartic coupling parameter and the modified coupling are of the same order $|\widetilde \lambda| \approx |\lambda|$, then for sufficiently small $f \ll |\lambda|^{-1}$ or very large occupancies $f \gg |\lambda|^{-1}$, such that $|\Pi_R| \ll 1$ or $|\Pi_R| \gg 1$, we have from (\ref{eq:leff})
\begin{align}
 \widetilde \lambda_{\mathrm{eff}}^2 \approx \widetilde \lambda^2\,, \qquad \qquad \widetilde \lambda_{\mathrm{eff}}^2 \sim \frac{\widetilde \lambda^2}{|\Pi_R|^2}\,,
\end{align}
respectively. In both cases only the square of the coupling enters the equation and there is no dependence on its sign. In the low-occupancy case, the perturbative Boltzmann equation (\ref{eq:perturbbolotz}) for the $\varphi^4$ theory with the prefactor for large $N$ and the replacement $\lambda \mapsto \widetilde \lambda$ is recovered. In contrast, in the other case of large occupancies, we have $\widetilde \lambda_{\mathrm{eff}}^2(t,\mathbf{p},\mathbf{q},\mathbf{r},\mathbf{l}) \ll \widetilde \lambda^2(t)$ and, therefore, the resummation generates an effectively weak coupling. Since $|\Pi_R|$ contains a factor of $\widetilde \lambda$, the modified coupling drops out of the entire Boltzmann equation, and it becomes identical to the resummed kinetic equation in $\varphi^4$ theory. In \cite{Berges:2015kfa, Orioli:2015dxa} self-similar solutions of the form (\ref{eq:self_similar}) have been studied for this regime. Under additional assumptions of particle number conservation and a nonrelativistic dispersion relation $\omega_{\mathbf{p}} \approx M +\mathbf{p}^2/2M$ for typical momenta, the values of the exponents $\alpha=3/2$, $\beta=1/2$ have been found. This solution describes the inverse particle cascade that we have observed also for our model in the repulsive and mean repulsive cases in Sec.~\ref{sec:dynamics_N1} with the same scaling exponents. 

However, if $|\Pi_R| \sim 1$, the sign of the modified coupling $\widetilde \lambda$ becomes important due to the denominators of (\ref{eq:leff}). In addition to this, the effective coupling has a pole at $\Pi_R \rightarrow -1$. If such a pole is encountered, the approximation based on the $1/N$ expansion may break down. 

To compare with our classical-statistical simulations for negative quartic coupling, we have shown the time evolution of the correspondingly computed modified couplings (\ref{eq:lmod}) in Figs.~\ref{fig:transition} and \ref{fig:transition28}, for the 1-, 2- and 8-component models. As can be seen, the regimes of mean repulsion and mean attraction are consistent with the modified coupling being positive or negative, respectively. When $\widetilde \lambda(t) \gtrsim |\lambda|$, the system enters the above mentioned self-similar regime after some time, which is rather well described by the vertex-resummed kinetic description. This does not happen numerically when $\widetilde \lambda(t) \sim -|\lambda|$, where number changing processes become important. These processes get also enhanced when $|\widetilde \lambda(t)| \ll |\lambda|$, and the kinetic formulation of the $1/N$ expansion up to NLO turns out to be insufficient to describe the essential aspects of the dynamics in these regimes.


\section{Conclusion}
\label{sec:conclusion}

We have studied the impact of attractive self-interactions on the nonequilibrium dynamics of relativistic scalar systems with large occupation numbers at low momenta. Focusing on $\mathcal{O}(N)$-symmetric field theories in $3+1$ dimensions with both quartic and sextic self-interactions, we compared the dynamics of the repulsive ($\lambda>0$) and attractive ($\lambda<0$) models.

In the repulsive theory with quartic and sextic interactions, we recover the same nonthermal attractor that has been observed already for the quartic theory. This observation is in line with the expectation that sextic couplings in $3+1$ dimensions represent irrelevant couplings in the renormalization group sense. The initially high population produces in this case a dual cascade in momentum space with self-similar dynamics at low and high momenta. The low momentum region involves a steep power law and leads to the formation of a Bose-Einstein condensate. Its scaling properties are insensitive to the number of field components.

This evolution is also observed if the classical quartic coupling is negative $\lambda < 0$, but the fluctuations are so large that the repulsive sextic interaction dominates and mean interactions are repulsive. In contrast, for smaller fluctuations in the mean attractive regime, the inverse cascade is absent. The particle annihilation rate is enhanced as compared to the repulsive case and no large spatial structures form, unless some nonvanishing conserved charge is present, which would forbid number changing processes and lead to the formation of $Q$-balls. In the absence of a charge, we have observed that enhanced particle loss can generate a transition between mean repulsion and mean attraction. This is accompanied by the collapse of the condensate and by a rapid decay of low momentum modes. We also observed that with an increasing number of field components, the system stays longer in the mean repulsive regime. 

From the $1/N$ expansion to NLO we have seen that the contribution from the sextic interaction, as well as from arbitrary interactions of the form $(\varphi_a \varphi_a)^{n}$ with $n \geq 3$, can be viewed as a time-dependent shift of a ``modified'' quartic coupling $\widetilde \lambda(t)$. This explains why the scaling solutions observed in the (mean) repulsive case are unaffected by the presence of higher-order self-interactions. Moreover, the observed transition between mean repulsion and mean attraction occurs approximately when this ``modified" quartic coupling changes its sign. 

An important application of our studies concerns the physics of dark matter axions and axion-like particles, which exhibit attractive self-interactions, as well as gravitational attractive interaction. Based on our analysis, we expect inelastic collisions to play an important role in the dynamics of such systems. Studying the impact of the phenomenologically more relevant gravitational interaction is left for future work.


\section {Acknowledgements}
We thank Asier Pi\~neiro Orioli, Tuomas Lappi and S\"oren Schlichting for helpful discussions. Some parts of this work are based on the master's thesis of A.C., that was supported by DAAD. This work is part of and supported by the DFG Collaborative Research Centre ``SFB 1225 (ISOQUANT)''. K.B.~is supported by the European Research Council under grant No.~ERC-2015-COG-681707. J.J.~gratefully acknowledges support by the DFG TransRegio research collaborative TR33 ``The Dark Universe''. Some of the numerical calculations were performed on the computational resource bwUniCluster, funded by the Ministry of Science, Research and the Arts Baden-W\"urttemberg and the Universities of the State of Baden-W\"urttemberg, within the framework program bwHPC.

 
\appendix


\section{Appendix A: Classical-statistical simulations}
\label{app:class_stat}

The basic idea of classical-statistical field theory is sampling over the initial conditions and evolving each realization according to the classical equation of motion. Observables are obtained by averaging over classical trajectories. In particular, the macroscopic field and the statistical propagator are given by
\begin{eqnarray}
\phi_a(x) & = & \langle \varphi_a(x) \rangle\,, \label{eq_field_propagator_1}\\
F_{ab}(x,y) & =& \langle \varphi_a(x) \varphi_b(y) \rangle - \phi_a(x)\phi_b(y)\,,
\label{eq_field_propagator_2}
\end{eqnarray}
where $\langle . \rangle$ denotes ensemble averages. Symmetries of the system like homogeneity and isotropy can also be used to provide additional averaging. 

Quantum fluctuations enter the classical-statistical approximation only through the initial conditions. In particular, Gaussian initial states for homogeneous systems, that are considered throughout this work, can be generated by
\begin{eqnarray}
\label{eq_gauss_distr_pi}
\nonumber
\varphi_a(t_0,\mathbf{x})&=&\phi_{0, a}+\int \frac{d^3\mathbf{p}}{(2\pi)^3} \sqrt{\frac{f_0(\mathbf{p})+1/2}{\omega_0(\mathbf{p})}}\;c_{a,\mathbf{p}}\, e^{i\mathbf{p}\mathbf{x}}\,,\label{eq_gauss_distr_fi} \\
\pi_a(t_0,\mathbf{x})&=&\pi_{0, a} +\int \frac{d^3\mathbf{p}}{(2\pi)^3} \sqrt{\Bigl(f_0(\mathbf{p})+1/2\Bigr) \,\omega_0(\mathbf{p})}\;\widetilde c_{a,\mathbf{p}}\, e^{i\mathbf{p}\mathbf{x}}\,.
\end{eqnarray}
Here $c_{a,\mathbf{p}}$ and $\widetilde{c}_{a,\mathbf{p}}$ are random Gaussian numbers multiplied by complex random phase factors, which satisfy $c_{a,\mathbf{p}}=c^{*}_{a,\mathbf{-p}}$ and $\langle c_{a,\mathbf{p}}c^*_{b,\mathbf{p'}}\rangle=\langle \widetilde c_{a,\mathbf{p}}\widetilde c^*_{b,\mathbf{p'}}\rangle=(2\pi)^3\delta_{ab}\delta(\mathbf{p}-\mathbf{p'})$, while all other correlations vanish. Equation (\ref{eq_gauss_distr_pi}) generates the following initial conditions for the one- and two-point functions (compare also to Sec.~\ref{sec:model} and footnotes therein)
\begin{equation}
\label{incond}
\begin{split}
F(t, t', \mathbf{p})\Big|_{t_0}= \frac{f_0(\mathbf{p})+1/2}{\omega_0(\mathbf{p})},  \: \: \partial_t \partial_{t'} F(t, t', \mathbf{p})\Big|_{t_0}= \Bigl(f_0(\mathbf{p})+1/2\Bigr)\, \omega_0(\mathbf{p}) \\
\partial_t F(t, t', \mathbf{p})\Big|_{t_0} = \partial_t' F(t, t', \mathbf{p})\Big|_{t_0} = 0, \: \: \: \: \: \phi(t_0)=\phi_0, \:\: \: \: \: \partial_t \phi(t_0)=\pi_0 \,. \\
\end{split}
\end{equation}
Higher order correlation functions in the case of a Gaussian state can be expressed in terms of one- and two-points correlators. In order to avoid quenches at initial time, we have initialized the modes with an effective mass, obtained by iteratively solving the gap equation 
\beq
M_0^2=m^2+\lambda \frac{N+2}{6N} \int_\mathbf{p} F(t_0,t_0,\mathbf{p}) + \lambda_6 \frac{(N+2)(N+4)}{5!N^2} \Bigl[ \int_\mathbf{p} F(t_0,t_0,\mathbf{p}) \Bigr]^2,
\eeq
with $\int_\mathbf{p} F(t_0,t_0,\mathbf{p})=V^{-1} \sum_\mathbf{p} \frac{f(t_0,\mathbf{p})+1/2}{\omega_0(\mathbf{p})}$ and initial dispersion relation $\omega_0(\mathbf{p}) = \sqrt{\mathbf{p}^2+M_0^2}$.

For the initial conditions (\ref{incond}) all $N(N-1)/2$ Noether charges given by Eq.~(\ref{eq:Noether}) vanish because of
\beq
\langle \hat Q_{ab} \rangle / V=\int_\mathbf{p} \Bigl[ \partial_{t'} F_{ab}(t,t',\mathbf{p}) - \partial_t F_{ab}(t,t',\mathbf{p}) \Bigr]\Big|_{t=t'} = 0\,.
\eeq
To generate a nonvanishing charge in the system, one can modify the conditions that the random numbers $c_a$ and $\tilde{c}_a$ from (\ref{eq_gauss_distr_pi}) have to satisfy. Especially, if one additionally imposes
\begin{equation}
\langle c_{a',\mathbf{p}}\tilde{c}_{b',\mathbf{p'}}\rangle=V \delta_{\mathbf{p,\:-p'}}\,
\end{equation}
for some $a'>b'$, then the charge density $\langle \hat Q_{a'b'} \rangle/V=\int_\mathbf{p}  (f_{\mathbf{p}}+\frac{1}{2})$ will be generated. 

Because of the large occupation numbers in the employed initial conditions (\ref{eq:box}), we omit the additional vacuum $1/2$ in the above formulas. While this does not change our results as long as typical occupation numbers are large $f \gg 1$, it allows a continuum extrapolation while avoiding the need to renormalize observables, which is in general not possible in classical-statistical theories \cite{Epelbaum:2014yja}.


\section{Appendix B: Transport equations from the 2PI $1/N$ expansion}
\label{app:transport_equ}

In this appendix, we describe how the effective kinetic equation (\ref{eq:kernel_resummed}) is derived starting from quantum field theory using the 2PI effective action. Although we consider the potential of (\ref{eq:potential}), the discussion can be easily generalized to arbitrary $\C{O}(N)$-symmetric polynomial potentials.

In the 2PI formalism the equations of motion for one- and two-point functions are given by the stationarity conditions (\ref{statconditions}). In the symmetric regime, the macroscopic field is zero and we are left with the equation for the propagator. The latter can be rewritten, taking into account the decomposition of (\ref{eq:expression_for_effaction}), in the following way \cite{Berges:2015kfa, Aarts:2002dj}
\beq
\label{gapequation}
G_{ab}^{-1}(x,y)=G_{0 \: ab}^{-1}(x,y)-\Sigma_{ab}(x,y; G),
\eeq
where 
\beq
\Sigma_{ab}(x,y;G)=2i\frac{\delta \Gamma_2[G]}{\delta G_{ab}(x,y)}
\label{selfenergy}
\eeq
is called proper self-energy and is the sum of all 1PI diagrams with two external lines.

After decomposing the full propagator $G_{ab}$ into spectral and statistical functions according to (\ref{separation}), and making a similar decomposition for the self-energy, with an additional separation of ``local'' and ``non-local'' parts,
\begin{eqnarray}
\label{eq:separationSelfEnergy1}
\Sigma_{ab}(x,y) \: = \: -i \: \Sigma_{ab}^{(0)}(x) \delta (x-y) \: + \: \bar{\Sigma}_{ab}(x,y), \\
\label{eq:separationSelfEnergy}
\bar{\Sigma}_{ab}(x,y) = \Sigma^{F}_{ab}(x,y) \: - \: \frac{i}{2} \Sigma^{\rho}_{ab}(x,y)\, \mathrm{sgn}_C(x^{0}-y^{0}),
\end{eqnarray}
the self-consistent equation (\ref{gapequation}) is replaced by an equivalent pair of coupled evolution equations for the statistical propagator and the spectral function \cite{Berges:2015kfa}:\begin{eqnarray}
\label{eq:coupledEvolution1}
[\Box_{x}\delta_{ac}+ M_{ac}^{2}(x)] F_{cb}(x,y) &=& -\int_{t_{0}}^{x_{0}} dz \Sigma_{ac}^{\rho}(x,z)F_{cb}(z,y) + \int_{t_{0}}^{y_{0}} dz \Sigma_{ac}^{F}(x,z)\rho_{cb}(z,y) \\
\label{eq:coupledEvolution2}
[\Box_{x}\delta_{ac}+ M_{ac}^{2}(x)] \rho_{cb}(x,y) &=& - \int_{y_{0}}^{x_{0}}dz \Sigma_{ac}^{\rho}(x,z)\rho_{cb}(z,y)
\end{eqnarray}
where $ \int_{t_{1}}^{t_{2}} dz\: = \: \int_{t_{1}}^{t_{2}} dz^{0} \int d^{d}z $ and (\ref{eq:propagator_cl}) has been used. An effective mass term $M^2_{ab}$ has been introduced, which in the symmetric regime is given by $M^2_{ab}(x)=m^2 \delta_{ab} + \Sigma^{(0)}_{ab}(x;G)$. These equations are called Kadanoff-Baym equations or 2PI equations of motion \cite{Kadanoff:1962}. If the expressions for the self-energies are known, one can prescribe the statistical propagator and its derivatives at the initial time (the equal-time spectral function is fixed by the commutation relations) and follow the time evolution of the two-point functions.

Transport equations are obtained from the 2PI equations of motion under several assumptions \cite{Berges:2015kfa, Berges:2005md}. One of them concerns the effective memory loss of the system about the initial conditions, which allows to take the limit $t_0 \rightarrow - \infty$. The next step is switching to relative and center coordinates
\begin{equation}
X^{\mu}=\frac{x^{\mu}+y^{\mu}}{2},\: \: \: \: s^{\mu}=x^{\mu}-y^{\mu}.
\end{equation}

Furthermore, a gradient expansion to the lowest order is employed, which means that only the lowest order terms in the number of derivatives with respect to the center coordinates  $X^{\mu}$ and powers of the relative coordinates $s^{\mu}$ are kept. Such description is expected to be suitable for a comparably smooth time evolution. 

Moreover, by virtue of $\mathcal{O}(N)$ rotations the propagators are considered to be diagonalized, so that $F_{ab}(x.y)=F(x,y) \delta _{ab}$ and $\rho_{ab}(x,y)=\rho(x,y) \delta_{ab}$. The two-point functions are Fourier transformed with respect to the relative coordinates
\begin{eqnarray}
F(X,p)  =  \int_s e^{ips}F(X+\frac{s}{2}, X-\frac{s}{2}), \:\:\:\:\:\: \: \: \:\:\:\:\: \widetilde{\rho}(X,p)  =  -i \int_s e^{ips}\rho(X+\frac{s}{2}, X-\frac{s}{2}),
\end{eqnarray}
where the factor $i$ is included in the definition of $\widetilde{\rho}$ to make it real-valued. Analogous transformations are done for the self-energies.

Taking into account all above mentioned approximations, the 2PI equations of motion can be recast into the following form \cite{Berges:2015kfa}
\begin{eqnarray}
\label{tpeq1}
2p^{\mu}\frac{\partial F(X,p)}{\partial X^{\mu}}&=&\widetilde{\Sigma}_\rho(X,p)F(X,p)-\Sigma_F(X,p)\widetilde{\rho}(X,p), \\
\label{tpeq2}
2p^{\mu}\frac{\partial\widetilde{\rho}(X,p)}{\partial X^{\mu}}&=&0.
\end{eqnarray}

Note that due to the gradient expansion to the lowest order the effective mass term is not present in these equations. For spatially homogenous systems the dependence on $X$ reduces to pure time dependence. Therefore, (\ref{tpeq2}) implies that $\widetilde{\rho}(p)$ does not depend on time. In other words, in this approximation the particle spectrum does not change with time. From (\ref{tpeq1}) the time evolution of the particle momentum distribution can be obtained. The latter can be defined according to \cite{Berges:2005md}
\begin{equation}
f(t,\mathbf{p})=\int_0^{\infty}\frac{dp^0}{2\pi}2p^0 \widetilde{\rho}(p) f(t,p)
\end{equation}
where $F(t,p)=\Bigl( f(t,p)+\frac{1}{2} \Bigr)\widetilde{\rho}(p)$.\footnote{This definition yields the same on-shell distribution function as our previous definition (\ref{eq:distr}) used for lattice simulations if the spectral function $\widetilde{\rho}(p)$ follows a $\delta$-like free-field form.} From (\ref{tpeq1}), the time evolution of the distribution function defined in that way is given by
\begin{equation}
\frac{\partial f(t,\mathbf{p})}{\partial t} =\int_0^\infty \frac{dp^0}{2\pi}\Bigr[ \widetilde{\Sigma}_\rho(t,p)F(t,p)-\Sigma_F(t,p)\widetilde{\rho}(p) \Bigl].
\label{grdExpansion}
\end{equation}
\bigskip

So far, the form of the potential has not been specified. Now we truncate the self-energies according to the truncation scheme of Sec.~\ref{sec:2PI_1N_exp}, that is based on a $1/N$ expansion to NLO. Taking into account (\ref{selfenergy}), as well as the expressions (\ref{eq:GammaLO}) and (\ref{eq:GammaNLO}), the LO contribution to the self-energy is given by
\beq
\Sigma_{ab}^{\mathrm{LO}}(x,y) = -i\Bigl( \frac{\lambda}{3! N}F_{cc}(x,x)+\frac{\lambda_6}{5!N^2}F_{cc}(x,x)F_{dd}(x,x) \Bigr) \delta_{ab}\delta(x-y),
\label{eq:selfEnergyLO}
\eeq
and the NLO contribution reads
\begin{eqnarray}
\nonumber
\nonumber \Sigma_{ab}^{\mathrm{NLO}}(x,y)=-i \Bigl(\frac{\lambda_6}{60N^2}\Bigr)  \int_{z,c}G_{mn}(x,z)G_{mn}(x,z)\widetilde B^{-1}(z,x;G)    \delta_{ab}\delta(x-y) - \\
- i\frac{\widetilde \lambda(y)}{3N}G_{ab}(x,y)\widetilde B^{-1}(x,y;G).
\label{eq:sEn}
\end{eqnarray}
In the second equation $\widetilde  B^{-1}$ sums the ``geometric series'':
\begin{eqnarray}
\nonumber
\widetilde B^{-1}(x,y;G)= \delta(x-y)-i\frac{\widetilde \lambda(x)}{6N}G_{ab}(x,y)G_{ab}(x,y)-\\
 -\: {\int_{z,C} }  \Bigl(  {\frac {\widetilde \lambda(x)}{6N}} \: G_{ab}(x,z)  G_{ab}(x,z) \Bigr) \Bigl( {\frac {\widetilde \lambda(z)}{6N}} G_{a^{\prime}b^{\prime}}(z,y)  G_{a^{\prime}b^{\prime}}(z,y) \Bigr)+...
\label{eq:B-1}
\end{eqnarray}

As can be seen, the LO contributes to $\Sigma_{ab}^{(0)}$ according to (\ref{eq:separationSelfEnergy1}), and therefore adds only to a coordinate-dependent mass shift to the free evolution, similar to the case of $\varphi^4$ theory \cite{Berges:2015kfa}. To separate the local and the non-local parts in $\Sigma^{\mathrm{NLO}}$ it is useful to split $B^{-1}(x,y)=\delta(x-y)-iI(x,y)$ and to rewrite (\ref{eq:sEn}) as
\begin{eqnarray}
\nonumber
 \Sigma_{ab}^{\mathrm{NLO}}(x,y)=-i \delta(x-y) \Biggl[   \delta_{ab}\frac{\lambda_6}{60N^2}  \Bigl(  G_{mn}(x,x)G_{mn}(x,x)- \\
-i\int_{z,C}G_{mn}(x,z) G_{mn}(x,z)I(z,x;G) \Bigr) +\frac{\widetilde \lambda(x)}{3N}G_{ab}(x,y)]   \Biggr] -\frac{\widetilde \lambda(y)}{3N}G_{ab}(x,y)I(x,y).
\end{eqnarray}

\begin{figure}
  \centering
    \includegraphics[scale=0.45]{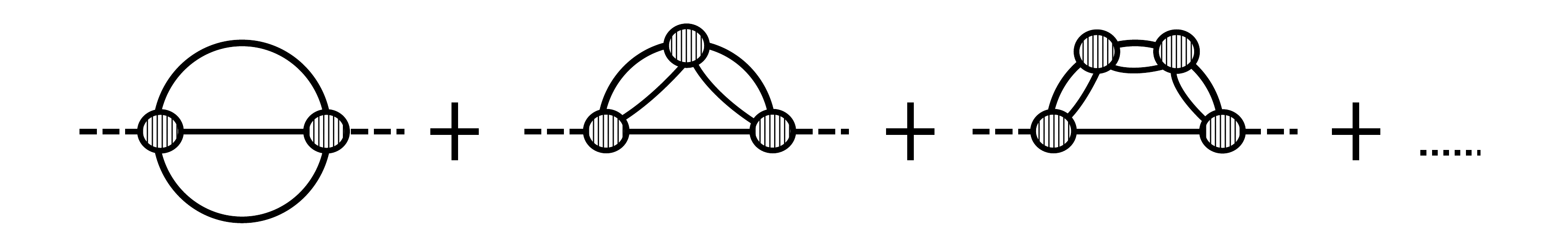}
\caption{The diagrammatic representation of $\bar{\Sigma}_{ab}^{\mathrm{NLO}}(x,y)$. The vertices with vertical lines correspond to the modified coupling $\widetilde \lambda$ from (\ref{eq:lmod}).}
\label{fig:self_energy}
\end{figure}

The first term with the $[..]$ brackets contributes to the effective mass, while the second term represents the non-local part of the self-energy and its diagrammatic representation is shown in Fig.~\ref{fig:self_energy}. Importantly, the expression of the non-local part $\bar{\Sigma}$ at NLO precisely coincides with the corresponding expression for the quartic theory \cite{Berges:2015kfa}, with the only difference being that $\lambda$ is replaced by $\widetilde \lambda(x)$. By introducing the following ``chain'' term 
\beq
\Pi(x,y) = \frac{\sqrt{\widetilde \lambda(x)\widetilde \lambda(y)}}{6N}G_{ab}(x,y)G_{ab}(x,y),
\eeq
the non-local part of the self-energy can be rewritten in an equivalent, but more symmetric way
\beq
\label{eq:nonlocalSelfEnergy}
\bar{\Sigma}_{ab}(x,y)=-\frac{\sqrt{\widetilde \lambda(x)\widetilde \lambda(y)}}{3N}G_{ab}(x,y)\C{I}(x,y),
\eeq
where $\C{I}(x,y)=\Pi(x,y)-i \int_{z,C} \C{I}(x,z) \Pi(z,y)$ sums the infinite series of chain diagrams and is connected to definitions above by $\C{I}(x,y) = I(x,y) \sqrt{\widetilde \lambda(y)/\,\widetilde\lambda(x)}$.

The non-local self-energy (\ref{eq:nonlocalSelfEnergy}) is further decomposed according to (\ref{eq:separationSelfEnergy}). This is done in a similar way as for the case of $\varphi^4$ theory \cite{Berges:2015kfa}. The components of the self-energy are given by
\begin{eqnarray}
\label{eq:similar1}
\Sigma^F_{ab}(x,y)&=&-\frac{\sqrt{\widetilde \lambda(x)\widetilde \lambda(y)}}{3N}\Bigl( F_{ab}(x,y)\C{I}_F(x,y)-\frac{1}{4}\rho_{ab}(x,y)\C{I}_\rho(x,y)  \Bigr), \\
\label{eq:similar2}
\Sigma^\rho_{ab}(x,y)&=&-\frac{\sqrt{\widetilde \lambda(x)\widetilde \lambda(y)}}{3N}\Bigl( F_{ab}(x,y)\C{I}_\rho(x,y) + \rho_{ab}(x,y)\C{I}_F(x,y)  \Bigr),
\end{eqnarray}
where the real-valued summation functions $\C{I}_F$ and $\C{I}_\rho$ are defined according to $\C{I}(x,y) = \C{I}^{F}(x,y) - \frac{i}{2} \C{I}^{\rho}(x,y)\, \mathrm{sgn}_C(x^{0}-y^{0})$ and satisfy the following relations:
\begin{eqnarray}
\C{I}_F(x,y) &=&  \Pi_F(x,y)-\int_{t^0}^{x^0} dz \C{I}_\rho(x,z)\Pi_F(z,y) +\int_{t^0}^{y^0} dz \C{I}_F(x,z) \Pi_\rho(z,y),\\
\C{I}_\rho(x,y) &=& \Pi_\rho(x,y) - \int_{y^0}^{x^0} dz\C{I}_\rho(x,z) \Pi_\rho(z,y).
\end{eqnarray}

The two components of the ``chain'' term are given by
\begin{eqnarray}
\label{eq:spComponents1}
\Pi_F(x,y) &=& \frac{\sqrt{\widetilde \lambda(x)\widetilde \lambda(y)}}{6N}\Bigl(F_{ab}(x,y)F_{ab}(x,y)-\frac{1}{4}\rho_{ab}(x,y)\rho_{ab}(x,y) \Bigr),\\
\label{eq:spComponents2}
\Pi_{\rho}(x,y) &=& \frac{\sqrt{\widetilde \lambda(x)\widetilde \lambda(y)}}{3N}F_{ab}(x,y)\rho_{ab}(x,y).
\end{eqnarray}

The last step is plugging the expressions (\ref{eq:similar1}-\ref{eq:spComponents2}) for the self-energies into the equation (\ref{grdExpansion}). We note that in the employed derivative expansion we have
\begin{equation}
\widetilde \lambda(x) = \widetilde \lambda \Bigl( X+\frac{s}{2} \Bigr) \approx \widetilde \lambda(X)+s^{\mu}\frac{\partial \widetilde \lambda(X)}{\partial X^{\mu}}\,, \qquad \sqrt{\widetilde \lambda(x)\widetilde \lambda(y)} \approx \widetilde \lambda(X)\,.
\end{equation}
Therefore, the only difference compared to the $\varphi^4$ theory \cite{Berges:2015kfa, Berges:2010ez} is the substitution $\lambda \rightarrow \widetilde \lambda(t)$, where $t$ is the central time coordinate. We state here the final result:
\begin {equation}
\begin{split}
C^{\mathrm{NLO}}[f](t,\mathbf{p})=\int d\Omega^{2\leftrightarrow 2}[f](t,p,l,q,r)[(f_p+1)(f_l+1)f_qf_r-f_pf_l(f_q+1)(f_r+1)] +\\
+\int d\Omega^{1\leftrightarrow 3}_{(a)}[f](t,p,l,q,r)[(f_p+1)(f_l+1)(f_q+1)f_r-f_pf_lf_q(f_r+1)] +\\
+\int d\Omega^{1\leftrightarrow 3}_{(b)}[f](t,p,l,q,r)[(f_p+1)f_lf_qf_r-f_p(f_l+1)(f_q+1)(f_r+1)] + \\
+\int d\Omega^{0\leftrightarrow 4}[f](t,p,l,q,r)[(f_p+1)(f_l+1)(f_q+1)(f_r+1)-f_pf_lf_qf_r]. \\
\end{split}
\label{BEqq}
\end{equation}

In the above expression $f_p=f(t,p)$ and the integration kernels are given by

\begin {equation}
\begin{split}
\int d\Omega^{2\leftrightarrow 2}[f](t,p,l,q,r)=\frac{\widetilde \lambda^2(t)}{18N}\int_0^\infty \frac{dp^0dl^0dq^0dr^0}{(2\pi)^{4-(d+1)}} \int_{\mathbf{lqr}}\delta(p+l-q-r) 
\widetilde\rho_p\widetilde\rho_l\widetilde\rho_q\widetilde\rho_r \times \\ \times [v_{eff}(t,p+l)+v_{eff}(t,p-q)+v_{eff}(t,p-r)], \\
\int d\Omega^{1\leftrightarrow 3}_{(a)}[f](t,p,l,q,r)=\frac{\widetilde \lambda^2(t)}{18N}\int_0^\infty \frac{dp^0dl^0dq^0dr^0}{(2\pi)^{4-(d+1)}} \int_{\mathbf{lqr}}\delta(p+l+q-r) 
\widetilde\rho_p\widetilde\rho_l\widetilde\rho_q\widetilde\rho_r \times \\ \times [v_{eff}(t,p+l)+v_{eff}(t,p+q)+v_{eff}(t,p-r)], \\
\int d\Omega^{1\leftrightarrow 3}_{(b)}[f](t,p,l,q,r)=\frac{\widetilde \lambda^2(t)}{18N}\int_0^\infty \frac{dp^0dl^0dq^0dr^0}{(2\pi)^{4-(d+1)}} \int_{\mathbf{lqr}}\delta(p-l-q-r) 
\widetilde\rho_p\widetilde\rho_l\widetilde\rho_q\widetilde\rho_r \times \\ \times  v_{eff}(t,p-l), \\
\int d\Omega^{0\leftrightarrow 4}[f](t,p,l,q,r)=\frac{\widetilde \lambda^2(t)}{18N}\int_0^\infty \frac{dp^0dl^0dq^0dr^0}{(2\pi)^{4-(d+1)}} \int_{\mathbf{lqr}}\delta(p+l+q+r) 
\widetilde\rho_p\widetilde\rho_l\widetilde\rho_q\widetilde\rho_r \times \\ \times v_{eff}(t,p+l). \\
\end{split}
\end{equation}
where $\int_{\mathbf{p}}=\int \frac{d^3\mathbf{p}}{(2\pi)^3}$ and $d=3$ for our case. The momentum-dependent four-vertex correction $v_{eff}$ is given by
\begin{eqnarray}
v_{eff}(t,p)=\frac{1}{|1+\Pi_R(t,p)|^2},
\label {veff}
\end{eqnarray}
where
\begin{equation}
\Pi_R(t,p)=\frac{\widetilde \lambda(t)}{3} \int_q F(t,p-q)G_R(q)
\label{retardprop}
\end{equation}
is the Fourier transformed retarded ``chain'' term, defined as $\Pi_R(x,y)=\Theta(x^0-y^0)\Pi_\rho(x,y)$.

Under an additional quasiparticle assumption, using the $\delta$-like free-field form of the spectral function
\begin{align}
 \widetilde{\rho}_0(p) = 2\pi\, {\rm sgn} (p^0)\, \delta\left( (p^0)^2 - \omega_{\mathbf{p}}^2 \right),
\end{align}
the terms corresponding to off-shell $1 \leftrightarrow 3$ and $0 \leftrightarrow 4$ processes vanish, and the term representing elastic $2 \leftrightarrow 2$ scatterings simplifies to (\ref{eq:kernel_resummed}).



\bibliographystyle{h-physrev5}
\bibliography{masterbib-new}

\end{document}